\documentclass[journal]{IEEEtran_modified}
\usepackage{cite}

\ifCLASSINFOpdf
  \usepackage[pdftex]{graphicx}
\else
\fi
\usepackage{amsmath}
\usepackage{amssymb}  
\usepackage{mathtools}

\ifCLASSOPTIONcompsoc
  \usepackage[caption=false,font=normalsize,labelfont=sf,textfont=sf]{subfig}
\else
  \usepackage[caption=false,font=footnotesize]{subfig}
\fi
\usepackage{booktabs}

\usepackage[nonumberlist, section, acronym, nomain]{glossaries}
\makeglossaries
\newacronym{RBPF}{RBPF}{Rao-Blackwellized particle filter}
\newacronym{CV}{CV}{constant velocity}
\newacronym{CA}{CA}{constant acceleration}
\newacronym{CTV}{CTV}{constant turn and velocity}
\newacronym{CTRV}{CTRV}{constant turn rate and velocity}
\newacronym{CTA}{CTA}{constant turn and acceleration}
\newacronym{CCA}{CCA}{constant curvature and acceleration}
\newacronym{EKF}{EKF}{extended Kalman filter}
\newacronym{UKF}{UKF}{unscented Kalman filter}
\newacronym{SIS}{SIS}{sequential importance sampling}
\newacronym{SIR}{SIR}{sequential importance resampling}
\newacronym{CSW}{CSW}{cumulative sum of weights}
\newacronym{MCMC}{MCMC}{Markov chain Monte Carlo}
\newacronym{SMC}{SMC}{sequential Monte Carlo}
\newacronym{ASIR}{ASIR}{auxiliary sampling importance resampling}
\newacronym{IMU}{IMU}{inertial measurement unit}
\newacronym{COG}{COG}{center of gravity}
\newacronym{ABD}{ABD}{adaptive breakpoint detector}
\newacronym{MMPF}{MMPF}{multiple model particle filter}
\newacronym{RDP}{RDP}{Ramer-Douglas-Peucker algorithm}
\newacronym{IMM}{IMM}{interacting multiple model}
\newacronym{RMSE}{RMSE}{root mean squared error}
\newacronym{DARPA}{DARPA}{Defense Advanced Research Projects Agency}
\newacronym{LMB}{LMB}{labeled multi-Bernoulli}
\newacronym{GM}{GM}{Gaussian mixture}
\newacronym{RFS}{RFS}{random finite set}
\newacronym{JIPDA}{JIPDA}{joint integrated probabilistic data association}
\newacronym{JPDA}{JPDA}{joint probabilistic data association}
\newacronym{PDA}{PDA}{probabilistic data association}
\newacronym{DGLMB}{$\delta$-GLMB}{$\delta$-generalized labeled multi-Bernoulli}
\newacronym{GLMB}{GLMB}{generalized labeled multi-Bernoulli}
\newacronym{PHD}{PHD}{probability hypothesis density}
\newacronym{CPHD}{CPHD}{cardinalized probability hypothesis density}
\newacronym{GP}{GP}{Gaussian process}
\newacronym{ADAS}{ADAS}{advanced driver assistance systems}
\newacronym{FISST}{FISST}{finite set statistics}
\newacronym{DGPS}{DGPS}{differential global positioning system}
\newacronym[plural=FOVs,firstplural=fields of view (FOVs)]{FOV}{FOV}{field of view}
\newacronym{MHT}{MHT}{multiple hypotheses tracking}
\newacronym{NN}{NN}{nearest neighbor}
\newacronym{EM}{EM}{expectation maximization}
\newacronym{RHM}{RHM}{random hypersurface model}
\newacronym{VGM}{VGM}{variational Gaussian mixture}
\newacronym{DBSCAN}{DBSCAN}{density-based spatial clustering of applications with noise}

\usepackage[capitalize]{cleveref}
\crefformat{equation}{(#2#1#3)}
\crefrangeformat{equation}{(#3#1#4) to~(#5#2#6)}
\crefmultiformat{equation}{(#2#1#3)}%
{ and~(#2#1#3)}{, (#2#1#3)}{ and~(#2#1#3)}

\usepackage{xcolor}
\usepackage{import}
\newcommand{\executeiffilenewer}[3]{%
    \ifnum\pdfstrcmp{\pdffilemoddate{#1}}%
    {\pdffilemoddate{#2}}>0%
    {\immediate\write18{#3}}\fi%
}

\DeclareMathOperator{\atantwo}{atan2}
\DeclareMathOperator{\St}{St}
\DeclareMathOperator{\Dir}{Dir}

\usepackage{mathtools}

\usepackage[keeplastbox]{flushend}

\usepackage{tikz}                      
\usepackage{pgfplots}                  
\pgfplotsset{compat=1.7}
\usepgfplotslibrary{external}		  
\usetikzlibrary{decorations.markings,arrows}
\tikzset{>=latex}
\graphicspath{{figures/}}
\newlength\figureheight
\newlength\figurewidth

\tikzset{every picture/.style={font issue=\footnotesize},
         font issue/.style={execute at begin picture={#1\selectfont}}
        }

\usepackage{siunitx}
\newcommand{\lState}{\mathbf{x}} 
\newcommand{\lmoState}{\mathbf{X}} 
\newcommand{\lmoPosterior}{{\boldsymbol{\pi}(\lmoState)}} 
\newcommand{\distinctLabelInd}{{\Delta(\lmoState)}} 
\newcommand{\labelProjectionX}{{\mathcal{L}(\lmoState)}} 
\newcommand{\lmb}{\gls{LMB}} 
\newcommand{\glmb}{\gls{GLMB}} 

\newcommand{\distinctLabel}[1]{{\Delta \left( #1 \right)}} 
\newcommand{\labelProjection}[1]{{\mathcal{L} \left( #1 \right)}} 
\newcommand{\inclusion}[2]{{1_{#1} \left( #2 \right)}} 
\newcommand{\lmoDensity}[1]{{\boldsymbol{\pi} \left( #1 \right)}} 

\newcommand{\useExternalFigures}{}

\definecolor{perulaBlue}{rgb}{0.0157,0.4,0.8157}%
\definecolor{perulaYellow}{rgb}{0.9647,0.9372,0.0824}%
\definecolor{perulaOrange}{rgb}{0.9960,0.7569,0.2274}%
\definecolor{uniGreen}{rgb}{0.33730,0.66670,0.10980}%
\definecolor{uniRed}{rgb}{0.63920,0.14900,0.21960}%

\usepackage{fancyhdr}
\begin{document}

\bstctlcite{IEEEexample:BSTcontrol}

%
\title{Tracking Multiple Vehicles Using a Variational Radar Model}

%
%

\author{Alexander Scheel
        and Klaus Dietmayer,~\IEEEmembership{Member,~IEEE}
\thanks{A. Scheel and K. Dietmayer are with the Institute of Measurement, Control, and Microtechnology, Ulm University, 89081 Ulm, Germany, e-mail: alexander.scheel@alumni.uni-ulm.de, klaus.dietmayer@uni-ulm.de}}%

\maketitle

\thispagestyle{fancy}

\begin{abstract}
High-resolution radar sensors are able to resolve multiple detections per object and therefore provide valuable information for vehicle environment perception. For instance, multiple detections allow to infer the size of an object or to more precisely measure the object's motion. Yet, the increased amount of data raises the demands on tracking modules: measurement models that are able to process multiple detections for an object are necessary and measurement-to-object associations become more complex. This paper presents a new variational radar model for tracking vehicles using radar detections and demonstrates how this model can be incorporated into a Random-Finite-Set-based multi-object filter. The measurement model is learned from actual data using variational Gaussian mixtures and avoids excessive manual engineering. In combination with the multi-object tracker, the entire process chain from the raw measurements to the resulting tracks is formulated probabilistically. The presented approach is evaluated on experimental data and it is demonstrated that the data-driven measurement model outperforms a manually designed model.
\end{abstract}


%
\IEEEpeerreviewmaketitle


\section{Introduction}\label{s:introduction}
\IEEEPARstart{R}{adar} sensors play an important role for vehicle environment perception due to their ability to directly measure the relative radial velocity of an object, their robustness to adverse weather conditions, and their low price. In particular, radar data is widely used to track other vehicles in an ego-vehicle's surrounding. Advances in automotive radar technology have led to increased sensor resolution and modern high-resolution radar is able to resolve multiple reflection centers of an object. Thus, each sensor may yield multiple measurements (i.e. detections) per object in a single scan. This additional data is valuable as it provides more information on the shape, extent, or motion of an object and facilitates tracking objects more precisely and in complex maneuvers.

However, tracking vehicles based on high-resolution radar data poses some challenges. First, one is faced with an extended object problem as the vehicle extent is---at least in the near field---not negligible in comparison to sensor resolution and multiple radar measurements from a vehicle need to be correctly processed to a single estimate. Thus, many classical filters, such as the Kalman filter, which suppose exactly one measurement per cycle, are not directly applicable. Radar data additionally exhibits some peculiarities which further complicate data processing: The detections may not always exhibit a clear shape and their number strongly depends on the sensor-to-object constellation. Also, the Doppler measurements introduce considerable ambiguity as they only provide the radial portion of an object's velocity and the superposition of forward motion and yaw rate causes different velocity vectors at different locations on the vehicle. Some Doppler measurements may even originate from rotating wheels and thus do not match the motion of the rigid body. In addition to data processing, the increased amount of detections further complicates the measurement-to-object association problem which is crucial in multi-object settings.

One solution to the extended object problem is to include preprocessing routines that reduce multiple measurements to a single meta-measurement. Several of such approaches have been proposed for radar-based tracking. These include clustering and extraction of reference points as in \cite{Dickmann.2015,Elfring.2016}, or fitting bounding boxes and L-shapes\cite{Roos.2016,Schlichenmaier.2017}, reflection center models \cite{Bordonaro.2015}, or velocity profiles \cite{Kellner.2013b,Kellner.2014,Kellner.2016} to the data. While preprocessing routines are oftentimes effective, computationally fast, and lead to clearly separable system architectures, they face difficulties if the data from a single time step is ambiguous and the correct meta-measurement cannot be easily extracted.

An alternative approach is to design extended object measurement models and filter algorithms which explicitly take all measurements into account. According to \cite{Granstrom.2017}, which provides an elaborate overview of extended object tracking, the approaches can be grouped into different modeling paradigms. The first paradigm models objects as a set of measurement sources with a specific spatial structure. An early version of this principle was presented in \cite{Salmond.1999} and variations of it have been applied to radar-based vehicle tracking in \cite{Darms.2009,Gunnarsson.2007,Hammarstrand.2012,Hammarstrand.2012b,Schuster.2014}.

A second variant of extended object models defines spatial distributions for the location of the measurement as initially proposed in \cite{Gilholm.2005b} and \cite{Gilholm.2005c}. A prominent example is the elliptical random matrix model \cite{Koch.2008} which has been extended for incorporating Doppler measurements in \cite{Schuster.2015}. Also, a polynomial object model for tracking stationary objects such as guard rails \cite{Lundquist.2010} and a Volcanormal density for modeling vehicles \cite{Broeit.2017} were proposed for radar applications.

The third paradigm is to choose a physics-based approach \cite{Granstrom.2017}. Although many of the aforementioned approaches (e.g. \cite{Hammarstrand.2012}) may as well be assigned to this category, it is used here to introduce \cite{Adam.2013} and \cite{Knill.2016} which use ray tracing to predict radar measurements. While \cite{Adam.2013} only considers the rear surface, the direct scattering model from \cite{Knill.2016} uses a full rectangular description of the vehicle which allows for tracking arbitrary maneuvers with varying aspect angles.

Lastly, some extended object models such as the random hypersurface model \cite{Baum.2014} or the Gaussian process model \cite{Wahlstroem.2015} use a parametric description of the object contour and estimate free form shapes. The Gaussian process approach has, for instance, been used for radar-based vehicle tracking in \cite{Michaelis.2017}.

One of the major advantages of extended object measurement models is that they work on the raw data directly. Thus, they use the entire available information and can resolve ambiguous situations by filtering over time. Still, some approaches such as the random matrix approach rely on restrictive assumptions that are not suitable for vehicle tracking. Others require a certain amount of modeling and implementation effort such as the approaches based on ray tracing or on sets of reflection centers. Yet, all share the drawback that expert knowledge and manual adaption are necessary for including a certain sensor effect such as the spurious measurements from rotating wheels.

Apart from tracking, sensor models are also important for sensor analysis and in simulation applications. Interestingly, there has been a recent development from radar models based on expert knowledge or physical calculations (e.g. \cite{Buhren.2006} and \cite{Schuler.2008}) towards data-driven approaches. For instance, \cite{Wheeler.2017} uses deep neural networks to simulate a radar power grid from an object list and a grid-based description of the environment. A statistical study on radar measurements from vehicles in dependence on the aspect angle was conducted in \cite{Berthold.2017}. In \cite{Hirsenkorn.2015} and \cite{Hirsenkorn.2016}, kernel density estimation methods are employed to learn a probabilistic measurement model for simulation.

In this paper, the idea of leaving the modeling task to machine learning tools is transferred to tracking. A variational radar model for vehicles is learned directly from actual radar detection data. Thus, the shortcomings of existing extended object measurement models are overcome: The engineering effort is diminished and different sensor effects are captured automatically. The process involves finding a conditional density function that relates the measurements and vehicle state. This is similar to the simulation model from \cite{Hirsenkorn.2016}. In this work, however, a \gls{VGM} approach \cite{Attias.1999,Attias.2000,Bishop.2013} is used. For automotive radar applications, \glspl{VGM} have, for example, previously been employed for batch estimation of maps in \cite{Lundgren.2016}. In contrast to kernel density estimation, they do not require storing all training data points and instead yield a compact analytical mixture density which can be easily incorporated into a tracking framework. As a Bayesian inference technique, \glspl{VGM} furthermore integrate nicely with tracking filters and facilitate an integral Bayesian view of the entire problem. Finally, \glspl{VGM} avoid well-known singularity issues of alternative \gls{EM} approaches and concurrently determine the number of required mixture components \cite{Bishop.2013}. To avoid the excessive manual labeling effort that oftentimes comes with machine learning, the training data set is automatically generated using a reference vehicle.

The variational radar model is additionally incorporated into a multi-object framework to track multiple vehicles and to tackle measurement-to-object associations, clutter measurements, and the fusion of radar data from multiple sensors in a principled way. In particular, an extended object \gls{LMB} filter \cite{Beard.2016} based on \gls{FISST} is chosen. \Gls{FISST} \cite{Mahler.2004,Mahler.2007} is a rather recent theoretical framework which provides a rigorous Bayesian formulation of the multi-object problem and mathematical tools for deriving different filter algorithms. Thus, it allows for a consistent probabilistic end-to-end formulation of the problem. Nonetheless, an adaption of other tracking approaches (multi-object or single object) to accommodate the variational radar model should be possible. See \cite{Granstrom.2017} for a more detailed overview of multi-object methods for tracking extended objects.

In the remainder of the paper, the tracking problem is first formulated in \cref{s:problemStatement}. The variational radar model and the multi-object measurement likelihood are then developed in \cref{s:measModel} and \cref{s:tracking} discusses the multi-object tracking approach. The application of the variational radar model to experimental data is shown in \cref{s:expRadarModel} and tracking results are evaluated in \cref{s:evaluation}. \Cref{s:conclusion} concludes the paper.


\section{Problem Formulation}\label{s:problemStatement}
\subsection{Vehicle and Measurement Representation}
\begin{figure}
\centering
	\def\svgwidth{.9\columnwidth}
	\vspace{0.2cm}
	\ifx\useExternalFigures\undefined
		\small%
    \executeiffilenewer{figures/schematic.svg}{figures/schematic.pdf}%
    {inkscape -z -D --file=figures/schematic.svg %
    --export-pdf=figures/schematic.pdf --export-latex}%
    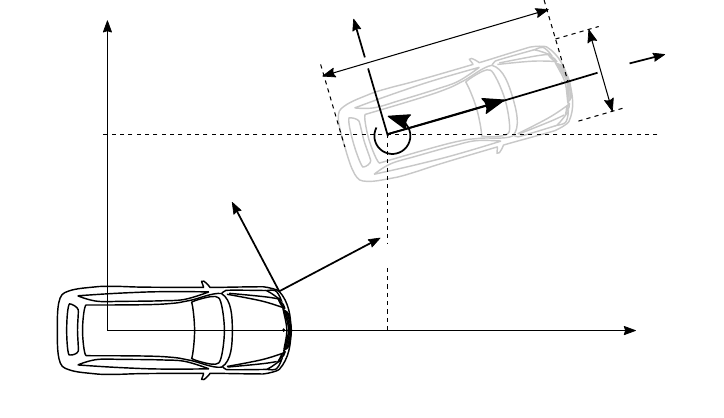%

	\else
		\small\input{figures/schematic.pdf_tex}
	\fi
	\caption{Schematic of the state vector, the radar measurements, and the vehicle (VC), sensor (SC), and object (OC) coordinate systems; adapted from \cite{Scheel.2016a}}
\label{fig:schematic}
\end{figure}

The goal is to recursively provide state estimates for all vehicles in the \gls{FOV} of the radar sensors based on the available measurements. As illustrated in \cref{fig:schematic}, each vehicle's state is described by the composed state vector $x_k = [\xi^T_k, \zeta^T_k]^T \in \mathbb{X}$ where $\mathbb{X}$ is the state space and $\xi_k$ and $\zeta_k$ are the kinematic and extent portion, respectively. Moreover, the subscript $k$ denotes the time step index. The kinematic state $\xi_k = [x_{R,k}, y_{R,k}, \varphi_k, v_k, \omega_k]^T$ combines the position of the rear axle center given by $x_{R,k}$ and $y_{R,k}$, the yaw angle $\varphi_k$, the vehicle speed $v_k$, and the yaw rate $\omega_k$. The extent portion $\zeta_k = [a_k, b_k]^T$ comprises the vehicle width $a_k$ and length $b_k$. The position of the rear axle is fixed at 77\% of the vehicle length as this value has empirically shown to be suitable for many vehicle types.

To identify the different objects and to extract trajectories over time, each state vector is augmented with a unique label $\ell \in \mathbb{L}$ from the label space $\mathbb{L}$. This yields the labeled state vector $\lState_k = [x_k^T, \ell]^T$. All present vehicles are combined in the multi-object state which is modeled as the \gls{RFS} $\lmoState_k = \{ \lState^{(1)}_k, \ldots, \lState^{(n)}_k\} \subset \mathbb{X} \times \mathbb{L}$ where the cardinality of the set $|\lmoState| = n$ equals the number of vehicles.

In each measurement cycle, a radar sensor provides a set of detections $Z_k = \{z^{(1)}_k, ..., z^{(m)}_k \} \subset \mathbb{Z}$ from the measurement space $\mathbb{Z}$ which either originate from actual vehicles, sensor noise, or other objects that are not relevant to the vehicle tracking task. Their number $m$ may change from cycle to cycle. Each detection $z_k = [d_k, \alpha_k, v_{D,k}]^T$ yields the measured range $d_k$, azimuth angle $\alpha_k$, and Doppler velocity $v_{D,k}$.

While the vehicle state is defined in the ego-vehicle coordinate system and the measurements are received in the sensor coordinate system using a polar representation, transformations to other coordinate system will be necessary. For instance, learning the vehicle model and computing the likelihood functions requires transforming the object states to the respective sensor coordinate system. Such transformations are indicated by the subscripts SC or OC for the sensor or object coordinate system when learning the variational model. To avoid cluttered notation, however, the subscripts are omitted in the measurement likelihood and filter update equations.

\subsection{The Multi-Object Bayes Filter}
The multi-object Bayes filter \cite{Mahler.2007} is used to recursively compute the posterior density of the multi-object state $\boldsymbol{\pi}_{k|k}(\lmoState_k | Z_{1:k})$. This density captures the uncertainty in both the number of set elements as well as their values and can hence be used to obtain estimates of the number of vehicles and their states. It is conditioned on all measurement sets from the first to the $k$-th time step as denoted by $Z_{1:k}$. As in the classical Bayes filter, the estimation procedure is split into a prediction and update step. In the prediction step, the prior multi-object density is computed using the Chapman-Kolmogorov equation
\begin{equation}
\begin{split}
&\boldsymbol{\pi}_{k | k-1}( \lmoState_k | Z_{1:k-1}) =\\
& \int \mathbf{f}_{k | k-1} (\lmoState_k | \lmoState_{k-1}) \boldsymbol{\pi}_{k-1 | k-1}(\lmoState_{k-1} | Z_{1:k-1}) \delta \lmoState_{k-1}, \\
\end{split}
\label{eq:moPrediction}
\end{equation}

where the multi-object transition density $\mathbf{f}_{k | k-1} (\lmoState_k | \lmoState_{k-1})$ governs the evolution of the multi-object state including object motion as well as appearance and disappearance. Information from new measurements is incorporated in the update step
\begin{equation}
\begin{split}
\boldsymbol{\pi}_{k | k} (\lmoState_k |& Z_{1:k}) = \\ 
&\frac
{g_k(Z_k | \lmoState_k) \boldsymbol{\pi}_{k | k-1} (\lmoState_k | Z_{1:k-1})}
{\int g_k(Z_k | \lmoState_k) \boldsymbol{\pi}_{k | k-1} (\lmoState_k | Z_{1:k-1}) \delta \lmoState_k}.
\end{split}
\label{eq:moUpdate}
\end{equation}
using the multi-object likelihood function $g_k(Z_k | \lmoState_k)$ which captures the measurement process and determines how likely the received measurements are for a specific multi-object state. As the computations involve set-valued random variables and their densities, the integrals in \cref{eq:moPrediction,eq:moUpdate} are set integrals as defined in \cite{Mahler.2007}. Note that the time subscript is dropped in the remainder of the paper to avoid cluttered notation. Prior quantities are indicated using the subscript $+$.

\section{Variational Radar Model}\label{s:measModel}
Before the multi-object filter from \cref{eq:moPrediction,eq:moUpdate} can be formulated in detail, the variational radar model and the multi-object likelihood function, which are required during filter update, are developed in this section. First, the basic concept of \gls{VGM}s is outlined. The approach is then applied to learning a model for a single vehicle. Finally, the model is incorporated into the multi-object likelihood.

\subsection{Variational Gaussian Mixtures}
\Glspl{VGM} for learning probabilistic models from data were initially presented in \cite{Attias.1999}. The basic assumption is that the data at hand is generated by an underlying Gaussian mixture model. However, the parameters of the model are unknown and the goal is thus to estimate the parameter values given the available data. This is done in a Bayesian fashion which involves computing posterior densities over the parameter values and allows for including a-priori knowledge about the parameters through prior densities. The estimated posterior parameter densities and the underlying Gaussian mixture then form a probabilistic model of the data which can be used to make predictions on future data points. In the following, the mathematical concepts are briefly outlined. The explanations closely follow \cite{Bishop.2013} to which the reader is referred to for a more detailed and accessible description.

Mathematically, the data for learning the model, the training data, is a set of $m$ data points $Z_D = \{z^{(1)}_{D}, \ldots, z^{(m)}_{D} \}$. Here, the letter $z$ is reused to emphasize that the training data is measured information even though it will have a different form than the presented radar measurements. The training data was created by a Gaussian mixture model with $c$ components. Each Gaussian distribution $\mathcal{N}(\cdot | \mu^{(j)}, \underline{H}_{(j)}^{-1})$ in the mixture is defined by its mean $\mu^{(j)}$, its precision matrix $\underline{H}_{(j)}$, and is assigned a mixing coefficient $w_j$ which measures the contribution of the $j$-th component to the density. For brevity, the mean vectors and precision matrices of all components are combined in the parameter sets $M$ and $H$, respectively, and the weights in the weight vector $w$. The latent variable $l^{(i)}$ is introduced to denote which component created the data point $z^{(i)}_D$. Hence, these latent variables are 1-of-K binary vectors where one of the elements $l^{(i)}_j$ is one and the remaining elements are zero. Again, all vectors are combined in the set of latent variables $L$. For given latent variables and parameter values, the likelihood of the training data is thus
\begin{equation}
p(Z_D | L, M, H) = \prod_{i = 1}^{m} \prod_{j = 1}^{c} \mathcal{N}(z^{(i)}_D | \mu^{(j)}, \underline{H}_{(j)}^{-1})^{l^{(i)}_j}.
\label{eq:gaussianMixture}
\end{equation}

Since the Bayesian treatment assumes the unknown latent variables and parameters to be random variables, the full probabilistic model is given by the joint distribution of the training data, latent variables, and parameters
\begin{equation}
\begin{split}
p(Z_D, L, & w, M, H) = \\
&p(Z_D | L, M, H) p(L | w) p(w) p(M | H) p(H).
\end{split}
\label{eq:varJoint}
\end{equation} 
The factorization follows from the Gaussian mixture structure and the Bayesian formulation. Its factors are the data likelihood, the distribution of the latent variables for given mixing coefficients
\begin{equation}
p(L | w) = \prod_{i = 1}^m \prod_{j = 1}^c w_j^{l^{(i)}_j},
\label{eq:gmLatent}
\end{equation}
and the prior distributions over the mixture model parameters $p(w)$, $p(M | H)$, and $p(H)$. These priors are modeled in conjugate forms to \cref{eq:gaussianMixture,eq:gmLatent}. The prior of the mixing coefficients is a Dirichlet distribution
\begin{equation}
p(w) = \Dir (w | \rho_0) = C(\rho_0) \prod_{j = 1}^c w_j^{\rho_0 - 1}
\label{eq:mixingPrior}
\end{equation}
with parameter $\rho_0$ and normalization constant $C(\rho_0)$. The prior of the mean vectors and precision matrices is a Gaussian-Wishart distribution with independent elements for each component. It is given by
\begin{equation}
\begin{split}
p(M, H) & = p(M|H) p(H) \\
= & \prod_{j=1}^c \mathcal{N}(\mu^{(j)} | \gamma_0, \beta_0^{-1} \underline{H}_{(j)}^{-1}) \mathcal{W}(\underline{H}_{(j)} | \underline{V}_0, \nu_0), \\
\end{split}
\label{eq:normalWishartPrior}
\end{equation}
with $\mathcal{W}(\cdot | \underline{V}_0, \nu_0)$ denoting a Wishart density and the parameters $\gamma_0$, $\beta_0$, $\underline{V}_0$, and $\nu_0$. Together with $\rho_0$, these are the hyperparameters of the model which govern the shape of the prior distributions and how informative they are. They are used to initialize all mixture components with the same value.

To compute the posterior densities over the latent variables and model parameters, a variational approach is used. It allows for an optimization-based approximation of the true posterior density and is based on maximizing the functional
\begin{equation}
\int q(\Phi) \ln \left( \frac{p(Z_D, \Phi)}{q(\Phi)} \right) \mathrm{d}\Phi.
\label{eq:varLowerBound}
\end{equation}
Here, the latent variables and model parameters were combined in $\Phi$ for brevity. The maximum of the functional occurs if the proposal distribution $q(\Phi)$ equals the true posterior distribution of the latent variables and model parameters $p(\Phi | Z_D)$ \cite{Bishop.2013}. Thus, the posterior distributions over parameters and latent variables are obtained by choosing a certain class of distributions for $q(\Phi)$ and maximizing \cref{eq:varLowerBound} with respect to $q(\Phi)$. For \glspl{VGM}, the factorized distribution
\begin{equation}
q(\Phi) = q(L, w, M, H) = q(L) q(w, M, H)
\end{equation}
is chosen. An optimal solution can then be found by iteratively maximizing \cref{eq:varLowerBound} with respect to $q(L)$ and $q(w, M, H)$. It can be shown that the optimal solution has the structure
\begin{equation}
q^\ast(L, w, M, H) = q^\ast(L) q^\ast(w) q^\ast(M | H) q^\ast(H)
\label{eq:optimalResult}
\end{equation}
where the different factors take the same form as the distributions from \cref{eq:gmLatent,eq:mixingPrior,eq:normalWishartPrior} with updated hyperparameters $\gamma^{(j)}$, $\beta_j$, $\underline{V}_{(j)}$, $\nu_j$, and $\rho_j$; see \cite{Bishop.2013} for the full equations. In contrast to the initial values of the hyperparameters, these values depend on the training data that is associated to the respective component and hence differ for each component.

To obtain the predictive density $p(\tilde{z}_D | Z_D)$ which measures how likely a new data point $\tilde{z}_D$ is, given the model that was obtained from the training data, the optimized distributions are inserted into \cref{eq:varJoint} and the latent variables as well as model parameters are marginalized by integration. This yields a mixture of Student's t~distributions \cite{Bishop.2013}
\begin{equation}
\begin{split}
p( &\tilde{z}_D | Z_D) = \\
& \frac{1}{\sum_{j = 1}^c \rho_j} \sum_{j = 1}^c \rho_j \St (\tilde{z}_D | \gamma^{(j)}, \underline{\tilde{H}}_{(j)}, \nu_j + 1 - |\tilde{z}_D|),
\end{split}
\label{eq:studentTMixture}
\end{equation}
where $\St (\cdot| \cdot, \cdot, \cdot)$ is a Student's t density, $\underline{\tilde{H}}_{(j)}$ is the precision matrix of the $j$-th component given by
\begin{equation}
\underline{\tilde{H}}_{(j)} = \frac{(\nu_j + 1 - |\tilde{z}_D|) \beta_j}{1 + \beta_j} \underline{V}_{(j)},
\end{equation}
and $|\tilde{z}_D|$ is the dimension of $\tilde{z}_D$. Note that the reason for obtaining a Student's~t mixture as predictive density instead of Gaussian mixture lies in the inclusion of the uncertainty in the parameter estimates.

\subsection{Learning a Variational Radar Model for Vehicles}\label{s:learningModel}
To obtain a vehicle measurement model, the \gls{VGM} approach is applied to actual radar measurements from vehicles and used to find a predictive density for radar detections given a particular vehicle state. Even though \glspl{VGM} are able to generalize to a certain extent, it is important to collect data samples from all relevant areas of the training data space to enable the \gls{VGM} to detect the structure and basic relationships in the data. For the presented state and measurement vectors, this would imply that data has to be collected in a ten-dimensional space. For instance, samples would be needed for vehicles of different size, with different poses, speeds, and yaw rates. Also, the complex relationship between Doppler measurements and object state as well as the representation of the measurements in polar coordinates may require many mixture components to be able to capture the nonlinearities.

To avoid these issues, the problem is simplified by applying dimension reduction. In particular, the measurements are transformed using the nonlinear transformation function
\begin{equation}
\begin{split}
z' = 
\begin{bmatrix}
z'_x \\
z'_y \\
z'_d \\
\end{bmatrix}
=  &
f_z (x_{SC}, z) \\
= &
\begin{bmatrix}
z_{x,OC} / b \\
z_{y,OC} / a \\
v_D - \left( \cos(\alpha) s_1 + \sin(\alpha) s_2 \right) \\
\end{bmatrix}, \\
\end{split}
\label{eq:zTransformationFunction}
\end{equation}
where $z_{x,OC}$ and $z_{y,OC}$ are the position of the radar detections in the object coordinate system,
\begin{equation}
s_1 = v \cos(\varphi_{SC}) + \omega y_{R,SC},
\label{eq:c1}
\end{equation}
and
\begin{equation}
s_2 = v \sin(\varphi_{SC}) - \omega x_{R,SC}.
\label{eq:c2}
\end{equation}
Thus, the position of all vehicle measurements is transformed to a normalized object coordinate system that is independent of the vehicle dimensions. This results in the coordinates $z'_x$ and $z'_y$. Additionally, the expected Doppler velocity is computed from the vehicle state using \cref{eq:c1,eq:c2}. It is subtracted from the measured Doppler velocity and the model therefore only learns the Doppler error $z'_d$. Here, information from the vehicle state is used and implicitly enters the measurement model. The object state itself is transformed using
\begin{equation}
x' = f_x (x_{SC}) = \varphi_{SC} - \atantwo (y_{R,SC}, x_{R,SC}).
\label{eq:xTransformationFunction}
\end{equation}
and is hence reduced to a single derived quantity which is approximately the aspect angle under which the sensor sees the vehicle. Concatenating $z'$ and $x'$ yields the training data representation $z_D = [z'^T, x']^T$.

Note that this manually designed dimension reduction requires some expert knowledge. Other techniques which include the distance to the vehicle as additional variable or which automatically detect a suitable representation (e.g. \cite{Tenenbaum.2000}) could have been used instead. However, the chosen variant can be interpreted to incorporate the most dominant and well-known properties where the additional effort to learn them does not appear to be beneficial. These properties are the basic Doppler measurement principle or the insight that the relative location of measurements will be approximately similar irrespective of the vehicle size or its position in the field of view and will mostly depend on the aspect angle.

By computing the predictive density \cref{eq:studentTMixture}, the \gls{VGM} model provides a joint distribution over the transformed measurements and state $p(z_D) = p(z', x')$, where the dependency on $Z_D$ is omitted for brevity. Then, the likelihood for the relative position of the measurements and the Doppler error for a given aspect angle $g_{z'}(z' | x')$ is obtained through
\begin{equation}
g_{z'}(z'|x') = \frac{p(z', x')}{p(x')},
\label{eq:conditionalDensity}
\end{equation}
where $p(x')$ is determined from marginalization. See \cite{Roth.2013} for the corresponding equations.

By using the \gls{VGM} technique, it is assumed that radar detections are generated by an underlying Gaussian mixture structure in which each measurement originates from one of the components. Intuitively, each mixture component can be interpreted to be a particular reflection center of the vehicle with associated position and measurement uncertainty. By including the aspect angle, the model not only learns the number and location but also the relevance of each reflection center for a particular line of sight. Yet, the mixture density is directly used as a spatial distribution model in this work to avoid an explicit association of detections to reflection centers during tracking.

\subsection{Multi-Object Likelihood Function}\label{s:moLikelihood}
So far, the presented approach allows learning a measurement model for a single vehicle which defines where radar detections are expected and how large the deviations from the expected Doppler velocity may be. For updating the multi-object state using \cref{eq:moUpdate}, however, the formulation of the entire multi-object likelihood $g(Z | \lmoState)$, which relates all measurements to all objects, is necessary.

\subsubsection{Detection-Type Likelihood}
To this end, the single-object model is incorporated into the multi-object likelihood function from \cite{Beard.2016} which is designed for detection-type measurements. It is based on several assumptions that have also been previously used in other extended object models (see e.g. \cite{Gilholm.2005c,Mahler.2009}). These assumptions are:
\begin{enumerate}
\item An object is detected with the probability of detection $p_D(x,\ell)$ or misdetected with the complementary probability $q_D(x, \ell) = 1 - p_D(x, \ell)$.
\item If an object is detected, it gives rise to a set of measurements $Z_O$ which follows the single object likelihood function $g(Z_O| x, \ell)$. The number of received measurements is Poisson distributed with expected value $\lambda_T$.
\item The measurement set $Z$ is a union of object and clutter measurement sets. The object measurement sets are independently generated by each object. 
\item The number of clutter measurements is Poisson distributed with expected value $\lambda_C$ and the values follow the density $p_C(z)$. Hence, they are distributed according to the Poisson \gls{RFS} \cite{Mahler.2007} $g_C$ with intensity function $\kappa(z) = \lambda_C p_C(z)$.
\end{enumerate}
Using these assumptions, the likelihood of obtaining a set of measurements from a given multi-object state is \cite{Beard.2016}
\begin{equation}
g(Z|\lmoState) = g_C(Z) \sum_{i=1}^{|\lmoState|+1} \sum_{\substack{\mathcal{U}(Z) \in \mathcal{P}_i(Z) \\ \theta \in \Theta(\mathcal{U}(Z))}} \left[ \psi_{\mathcal{U}(Z)}(\cdot | \theta)\right]^\lmoState
\label{eq:moLikelihood}
\end{equation}
with
\begin{equation}
\psi_{\mathcal{U}(Z)}(x, \ell | \theta) = \left\{
\begin{array}{ll}
\frac{p_D(x,\ell) g(\mathcal{U}_{\theta(\ell)}(Z)|x,\ell)}{[\kappa(\cdot)]^{\mathcal{U}_{\theta(\ell)}(Z)}}, & \theta(\ell) > 0 \\
q_D(x,\ell), & \theta(\ell) = 0 \\
\end{array}
\right.,
\label{eq:psi}
\end{equation}
\begin{equation}
g_C(Z) = e^{-\lambda_C} [\kappa(\cdot)]^Z.
\label{eq:clutter}
\end{equation}
Here, the short notation
\begin{equation}
h^X \triangleq \prod_{x \in X} h(x), \; \; h^\emptyset = 1,
\label{eq:multiObjectExponential}
\end{equation}
is used to denote products of a real-valued function $h(\cdot)$ applied to all elements of a set. In a nutshell, the function evaluates the different possibilities of how the measurements could be composed and computes their likelihood. For this purpose, the two sums in \cref{eq:moLikelihood} are used to evaluate different partitions $\mathcal{U}(Z)$ of the measurement set and different association mappings $\theta$. $\mathcal{P}_i(Z)$ denotes the set of all partitions that contain $i$ mutually exclusive clusters. Each association mapping $\theta : \labelProjection{\lmoState} \rightarrow \{0, 1, \ldots, |\mathcal{U}(Z)| \}$ assigns the labels from the multi-object state to the clusters in a partition. The labels are retrieved using the label projection function $\labelProjection{\lmoState} = \{ \ell  \mid [x^T, \ell]^T \in \lmoState\}$.

A cluster in a partition may only be assigned to one track, i.e. $\theta(\ell) = \theta(\ell') > 0$ implies $\ell = \ell'$ while several tracks may be assigned to the index 0 which stands for a misdetection. $\Theta(\mathcal{U}(Z))$ is the space of all possible association mappings and the cluster assigned to track $\ell$ is identified by $\mathcal{U}_{\theta(\ell)}(Z)$.

For the case $\theta(\ell) > 0$, \cref{eq:psi} computes the single object likelihood
\begin{equation}
g(Z_O | x, \ell) = e^{-\lambda_T} [\lambda_T g_z(\cdot | x)]^{Z_O},
\label{eq:soLikelihood}
\end{equation}
where $Z_O = \mathcal{U}_{\theta(\ell)}(Z)$ for a specific track-to-cluster association, and cancels the measurements from the overall clutter term $g_C(Z)$. Reformulating the ratio
\begin{equation}
\frac{g(Z_O|x,\ell)}{[\kappa]^{Z_O}}
=
\frac{e^{-\lambda_T} \lambda_T^{|Z_O|} }{\lambda_C^{|Z_O|}} \prod_{z \in Z_O} \frac{g_z(z|x)}{p_C(z)}
\label{eq:reformRatio}
\end{equation}
from \cref{eq:psi} separates it into a factor which considers the number of measurements and a factor which compares how well measurements fit to the object and clutter likelihoods. 

\subsubsection{Incorporating the Variational Model}
The multi-object likelihood is a density over an \gls{RFS} that is a subset of the measurement space $\mathbb{Z}$. Hence, the object and clutter likelihoods are densities with $\mathbb{Z}$ as sample space. In contrast, the conditional density \cref{eq:conditionalDensity} from the variational radar model is a density over a normalized space where the scaling depends on the object state. Simply inserting \cref{eq:conditionalDensity} into \cref{eq:soLikelihood} is thus mathematically incorrect and would prohibit a meaningful comparison between different tracks as well as clutter.

Yet, the identity
\begin{equation}
\begin{split}
\frac{g_{z'}(z'|x)}{p_C(z')}
= &
\frac{g_z (f_z^{-1}(z', x) | x ) \left\vert \frac{\partial f_z^{-1}(z',x)}{\partial z'_x \partial z'_y \partial z'_d} \right\vert}
{p_C ( f_z^{-1}(z', x) ) \left\vert \frac{\partial f_z^{-1}(z',x)}{\partial z'_x \partial z'_y \partial z'_d} \right\vert} \\
= &
\frac{g_z ( f_z^{-1}(z',x) | x )}
{p_C ( f_z^{-1}(z',x) )}
=
\frac{g_z\left( z | x \right)}
{p_C\left( z \right)}
\end{split}
\label{eq:ratio}
\end{equation}
states that the ratio between the object and clutter likelihood remains identical if both likelihoods are transformed using the same transformation function. Thus, it can be used to replace the likelihood ratio from \cref{eq:reformRatio} with a new ratio between the conditional density from the variational radar model and the transformed clutter density. The identity follows from computing the distributions of $z'$ as derived distributions from $z$; see e.g. \cite{Gamarnik.2008}. Note that the inverse of the transformation function $f_z^{-1}(z', x)$, which transforms the measurements from a Cartesian representation back to the original polar measurement space, exists. Yet, it is not defined at the location of the sensor origin. As this pathological case is not relevant in practical scenarios, it is neglected here. Also, $g_{z'}(z' | x) = g_{z'}(z' | x')$ as the information of $x'$ is fully contained in $x$.

One way to obtain the clutter density for the transformed measurements $p_C(z')$ is to fully transform the original clutter density over the polar measurement space. Here, however, an alternative approach is chosen: It is assumed that clutter is uniformly distributed over the Cartesian sensor coordinate system. Moreover, a mixture between a uniform density and a Gaussian distribution is used for modeling the Doppler values. The Gaussian distribution is centered at the Doppler value of stationary objects and emphasizes that they are the most frequent clutter source. Using the Cartesian representation, the transformation of the density to the space of $z'$ is considerably simplified and mostly involves scaling by the factor $a \cdot b$ to account for the vehicle size. The corresponding clutter density in the original measurement space which is required in \cref{eq:moLikelihood} could be determined by transformation. Yet, this factor cancels in the update step and is not required; see \cref{s:update}.

In summary, the ratio \cref{eq:ratio}, which is inserted into the multi-object likelihood, is computed by first transforming the current measurements and state and then evaluating the densities. Note that information from the state enters through both steps.

\section{Multi-Object Tracking}\label{s:tracking}
\begin{figure}
\centering
	\def\svgwidth{.9\columnwidth}
	\vspace{0.2cm} 
	\ifx\useExternalFigures\undefined
		\footnotesize%
    \executeiffilenewer{figures/systemSchematic2.svg}{figures/systemSchematic2.pdf}%
    {inkscape -z -D --file=figures/systemSchematic2.svg %
    --export-pdf=figures/systemSchematic2.pdf --export-latex}%
    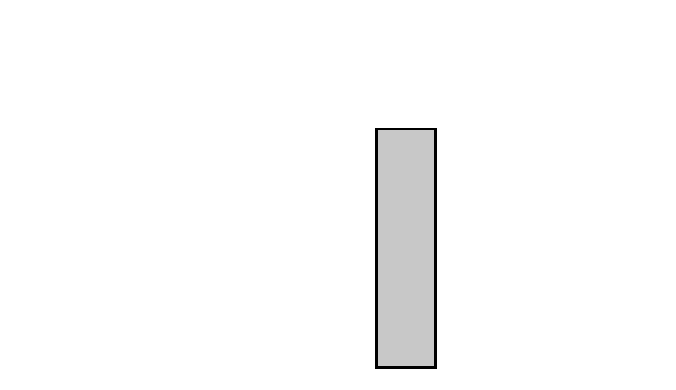%

	\else
		\footnotesize\input{figures/systemSchematic2.pdf_tex}
	\fi
	\caption{Overview of the filtering procedure}
\label{fig:systemSchematic}
\end{figure}

For tracking multiple vehicles, the multi-object measurement likelihood from the previous section is used in an extended object \gls{LMB} filter \cite{Beard.2016}. This filter has, for instance, also been used in \cite{Scheel.2016a} in conjunction with the direct scattering model. By modeling the multi-object state using \gls{LMB} and \gls{GLMB} distributions \cite{Vo.2013}, it facilitates an analytical solution to \cref{eq:moPrediction,eq:moUpdate}. Please refer to \cite{Beard.2016} for a detailed description including pseudo code. In this paper, the original version is slightly modified to avoid overlapping objects as initially proposed in \cite{Scheel.2016b}. A schematic overview of the filtering procedure is shown in \cref{fig:systemSchematic}.

\subsection{Initialization and Prediction}
At the end of the last filter recursion and before prediction, the distribution over the current multi-object state is represented using an \gls{LMB} distribution. It consists of several independent object hypotheses which are described by the existence probability $r^{(\ell)}$ and the single object state density $p(x, \ell)$. The labels of all present object hypotheses define the label space $\mathbb{L}$. The multi-object density is thus given by
\begin{equation}
\lmoPosterior = \distinctLabelInd w(\labelProjectionX) [p(\cdot)]^\lmoState,
\label{eq:lmbDensity}
\end{equation}
where
\begin{equation}
w(I) = \prod_{i \in \mathbb{L}} \left( 1 - r^{(i)} \right) \prod_{\ell \in I} \frac{1_\mathbb{L}(\ell) r^{(\ell)}}{1 - r^{(\ell)}} 
\label{eq:lmbWeight}
\end{equation}
is the probability that all tracks in the multi-object state $\lmoState$ exist and the remaining hypotheses do not. The inclusion function $1_\mathbb{L}(\ell)$ ensures that only labels from existing hypotheses are used and is 1 if and only if $\ell \in \mathbb{L}$. Moreover the distinct label indicator $\distinctLabel{\lmoState} = \delta(|\labelProjection{\lmoState}| - |\lmoState|)$ where $\delta(\cdot)$ denotes the Kronecker-delta function is used to ensure that each object in a labeled set has a unique label. The cardinality distribution can be obtained by marginalizing over the states which yields a Poisson binomial distribution. The expected value of this distribution serves as cardinality estimate and is the sum over the existence probabilities of all hypotheses.

In the track initialization stage, new track hypotheses are generated for measurements that have not considerably contributed to updating existing tracks and exhibit a significant Doppler velocity. The new hypotheses are labeled with new labels from label space $\mathbb{B}$ and are assigned an initial existence probability $r^{(\ell)}_B$ as well as a prior state density $p_B(x, \ell)$. Afterwards, they are appended to the existing hypotheses which yields the new and augmented label space $\mathbb{L}_+ = \mathbb{L} \cup \mathbb{B}$.

In the first prediction step, the existing and new tracks are predicted using the standard multi-object transition model. Each object survives to the next time step with a probability of persistence $p_S(x, \ell)$ or disappears with complementary probability. If an object survives, its states evolve according to the single object transition density $f_+(x_+|x, \ell)$. Hence \cite{Beard.2016},
\begin{align}
r_+^{(\ell)} & = \eta(\ell)r^{(\ell)} \label{eq:lmbPriorComponents1}, \\
p_+(x_+,\ell) & = \frac{\int p_S(x, \ell)  f_+(x_+ | x, \ell) p(x, \ell) \mathrm{d} x}{\eta(\ell)} \label{eq:lmbPriorComponents2}, \\
\eta(\ell) & = \iint p_S(x, \ell)  f_+(x_+ | x, \ell) p(x, \ell) \mathrm{d}x \mathrm{d}x_+.
\label{eq:lmbPriorComponents}
\end{align}
Here, $x_+$ is the predicted state, $r^{(\ell)}_+$ the predicted existence probability and $p_+(x_+, \ell)$ the predicted state density of hypothesis $\ell$. The new densities and existence probabilities then constitute the parameters of the prior \gls{LMB} distribution from the first prediction step. For the application to vehicle tracking, $f_+(x_+|x, \ell)$ consists of a \gls{CTRV} motion model \cite{Schubert.2008} with additive noise for the kinematic state and pseudo noise is added to the extent portion. 

Subsequently, a second prediction step eliminates hypotheses with overlapping objects by conditioning the predicted multi-object density on the event of physical feasibility $\mathcal{F}$. This yields \cite{Scheel.2016b}
\begin{equation}
\boldsymbol{\pi}_+(\lmoState_+ | \mathcal{F}) = \distinctLabel{\lmoState_+} w_+(\labelProjection{\lmoState_+}) [p_+(\cdot)]^{\lmoState_+},
\label{eq:lmbPredDensity1}
\end{equation}
with
\begin{equation}
w_+(I) = \frac
{p(\mathcal{F}|I) \tilde{w}_+(I) }
{\sum_{J \subseteq \mathbb{L}_+} p(\mathcal{F}|J) \tilde{w}_+(J)},
\label{eq:lmbPredDensity2}
\end{equation}
and where $\tilde{w}_+(I)$ is obtained by inserting the existence probabilities from the first prediction step \cref{eq:lmbPriorComponents1} into \cref{eq:lmbWeight}. The likelihood for physical feasibility $p(\mathcal{F}|I)$ is chosen to be 1 if and only if none of the objects in the label set $I$ overlap. Here, the predicted mean values of the vehicle positions and extents are used to determine possible overlaps.

Yet, the prior multi-object density from \cref{eq:lmbPredDensity1,eq:lmbPredDensity2} is not in \gls{LMB} form and objects are not independent anymore. That is, the weight does no longer factorize over the set elements. Instead, the multi-object prior is now a variant of the more general \gls{GLMB} distribution which allows for arbitrary weights and superposition of several multi-object hypotheses \cite{Vo.2013}.

\subsection{Update}\label{s:update}
Substituting the multi-object likelihood equations from \cref{s:moLikelihood} and the multi-object prior from the prediction into \cref{eq:moUpdate} yields the parameters of the posterior multi-object distribution (cf. \cite{Beard.2016})
\begin{equation}
\begin{split}
\lmoDensity{\lmoState|Z} =& \distinctLabel{\lmoState} \sum_{i=1}^{|\lmoState|+1} \sum_{\substack{\mathcal{U}(Z) \in \mathcal{P}_i(Z) \\ \theta \in \Theta(\mathcal{U}(Z))}} w_{\mathcal{U}(Z)}(\labelProjection{\lmoState} | \theta) \\
& \times \left[ p(\cdot|\mathcal{U}(Z), \theta) \right]^\lmoState \\
\end{split}
\label{eq:posteriorDensity}
\end{equation}
with
\begin{equation}
w_{\mathcal{U}(Z)}(I | \theta) = \frac{w_+(I) \left[ \eta_{\mathcal{U}(Z)}(\cdot | \theta) \right]^I}
{\sum \limits_{J \subseteq \mathbb{L}} \sum \limits_{i=1}^{|J|+1} \smashoperator[r]{\sum \limits_{\substack{\mathcal{U}(Z) \in \mathcal{P}_i(Z) \\ \theta \in \Theta(\mathcal{U}(Z))}}} w_+(J) \left[ \eta_{\mathcal{U}(Z)}(\cdot | \theta) \right]^J},
\label{eq:posteriorWeights}
\end{equation}
\begin{equation}
p(x,\ell|\mathcal{U}(Z), \theta) = \frac{p_+(x_+,\ell) \psi_{\mathcal{U}(Z)}(x_+,\ell | \theta)}
{\eta_{\mathcal{U}(Z)}(\ell | \theta)},
\label{eq:soUpdate}
\end{equation}
and
\begin{equation}
\eta_{\mathcal{U}(Z)}(\ell | \theta) = \int p_+(x_+,\ell) \psi_{\mathcal{U}(Z)}(x_+,\ell | \theta) \mathrm{d}x_+.
\end{equation}
Again, the distribution is in \gls{GLMB} form. Yet, the update step introduces additional dependencies among objects which arise from the claim that each cluster in a measurement partition may only be assigned to one object. As observable from the sums in \cref{eq:posteriorDensity}, the posterior multi-object distribution is hence composed of several hypotheses that have been updated using different partitioning and clustering possibilities.

\subsection{Approximation}
To avoid a steady increase of multi-object hypotheses in the \glmb\ posterior over time, the posterior \glmb\ density is approximated by an \lmb\ density at the end of each filter recursion. This procedure has been proposed by \cite{Reuter.2014} and results in a posterior \lmb\ density with parameters \cite{Beard.2016}
\begin{equation}
r^{(\ell)} = \sum_{I \subseteq \mathbb{L}_+} \sum_{i=1}^{|I|+1} \sum_{\substack{\mathcal{U}(Z) \in \mathcal{P}_i(Z) \\ \theta \in \Theta(\mathcal{U}(Z))}} w_{\mathcal{U}(Z)}(I | \theta) \inclusion{L}{\ell},
\end{equation}
and
\begin{equation}
\begin{split}
p(x, \ell) =& \frac{1}{r^{(\ell)}} \sum_{I \subseteq \mathbb{L}_+} \sum_{i=1}^{|I|+1} \sum_{\substack{\mathcal{U}(Z) \in \mathcal{P}_i(Z) \\ \theta \in \Theta(\mathcal{U}(Z))}} w_{\mathcal{U}(Z)}(I | \theta) \inclusion{L}{\ell} \\
& \times p(x, \ell | \mathcal{U}(Z), \theta).
\end{split}
\end{equation}
Vehicle state estimates are then extracted from this result.

\subsection{Estimating the Single Object Densities}
The extended \lmb\ filter internally holds and processes the state densities of the different object hypotheses. In particular, \cref{eq:lmbPriorComponents2,eq:soUpdate} predict the single object state and update it with the associated measurements, respectively. To solve these equations, standard Bayesian filtering techniques can be applied. This work uses a particle filter approach as both the transition density and the measurement model are nonlinear.

To reduce the amount of required particles, a simplified approach which is based on the \gls{RBPF} technique \cite{Murphy.2001} is applied. Only the kinematic portion $\xi$ is fully represented by particles while the estimation of the extent portion $\zeta$  is approximated by employing discrete distributions. At the beginning of the filter procedure, each particle holds a single hypothesis for the vehicle extent. During prediction, a discrete transition density is applied to each particle. It creates new extent hypotheses by varying the width and length. Thus, a discrete distribution with up to nine elements is generated. The likelihood is evaluated for all extent hypotheses and the resulting posterior extent distribution of each particle is again reduced to a single extent hypothesis by computing its mean. Note, however, that this step discards information about the extent estimate and that the particles hence do not capture the full extent uncertainty. Yet, the entire procedure introduces a local search for best fitting extent and allows an easy adaption of each particle's extent estimate.

\section{Radar Model from Experimental Data}\label{s:expRadarModel}
\newcommand{\nDSRuns}{20}
\newcommand{\correctCardDS}{66.8\%}
\newcommand{\underestCardDS}{15.7\%}
\newcommand{\overestCardDS}{17.5\%}
\newcommand{\cardErrorLateInitDS}{6.2\%}
\newcommand{\nVMRuns}{20}
\newcommand{\correctCardVM}{73.8\%}
\newcommand{\underestCardVM}{11.0\%}
\newcommand{\overestCardVM}{15.2\%}
\newcommand{\cardErrorLateInitVM}{6.9\%}
\newcommand{\tIntVMx}{\SI{0.21}{m}}
\newcommand{\tIntVMy}{\SI{0.52}{m}}
\newcommand{\tIntVMPhi}{\SI{4.3}{\degree}}
\newcommand{\tIntVMa}{\SI{0.40}{m}}
\newcommand{\tIntVMb}{\SI{0.54}{m}}
\newcommand{\tIntVMTrackAvail}{95.1\%}
\newcommand{\nRadarRuns}{20}

Now that the measurement model and multi-object filter are formulated, they are applied to experimental radar data. This section first describes the process of learning a variational radar model for vehicles. As a supplement, the resulting model is made available online\footnote{The variational radar model is available at\\https://github.com/A-Scheel/Variational-Radar-Model}. The application to vehicle tracking is then demonstrated in the following section.

\subsection{Experimental Set-Up and Data Set}
To generate the measurement data, two vehicles were used. The ego-vehicle is equipped with four short-range radar sensors that are mounted in the corners of the front and rear bumper. The sensors have an opening angle of about \ang{170}, a range of \SI{43}{m}, and the sensor axes are rotated by \ang{45} with respect to the vehicle axis. Thus, an almost complete \ang{360} coverage of the close-up range is given. All sensors run at a frequency of \SI{20}{Hz} and are not synchronized among themselves. Apart from the radar sensors, an IBEO Lux lidar, which serves as reference sensor, is mounted in the center of the front bumper. The second vehicle, a Mercedes E-Class station wagon (S212), serves as target vehicle. Both vehicles are equipped with a GeneSys ADMA which combines a precise \gls{DGPS} and \gls{IMU}. It provides the pose of the vehicles in a global coordinate system and the object motion. This allows computing the ground truth position of the target vehicle in both the ego-vehicle coordinate system and the four sensor coordinate systems.

The measurement data was collected on a closed test site and on public roads. It includes typical longitudinal and cross traffic situations as well as artificial maneuvers which were designed to achieve a good coverage of the measurement and state space. These maneuvers, for instance, include circling the stationary ego-vehicle in different distances, driving small circles in different parts of the \gls{FOV}, or driving straight lines at different distances and angles.

In case of a stationary vehicle, measuring the orientation in global coordinates is challenging for the \gls{DGPS}/\gls{IMU} system. From eye inspection, a mismatch between the vehicle ground truth and the laser measurements was observable in some sequences. In these cases, the orientation error was manually corrected using the precise measurements from the lidar.

The data set was then generated from the recorded measurements by computing the ground truth position of the target vehicle in sensor coordinates and determining the measurements that originate from the vehicle by gating. That is, only radar detections in a bounding box that exceeds the actual vehicle dimensions by \SI{0.5}{m} in all directions were paired with the respective ground truth vehicle state. The remaining clutter and measurements from other traffic participants were discarded. Subsequently, the transformation functions \cref{eq:zTransformationFunction,eq:xTransformationFunction} were applied to the extracted detection and vehicle state pairs.

\begin{figure}
\centering
\setlength\figurewidth{7.5cm}
\subfloat[Measurement positions in normalized object coordinates]{
	\ifx\useExternalFigures\undefined
    \tikzsetnextfilename{ext_figures/trainingDataZxZy}%
    \input{figures/trainingDataZxZy.tikz}%

	\else
		\includegraphics{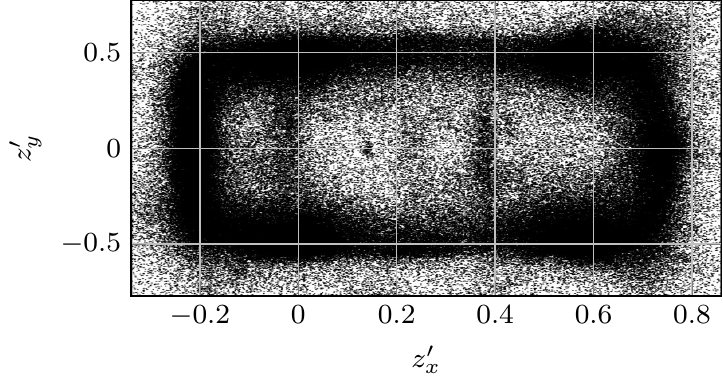}
	\fi
	\label{fig:trainingDataZxZy}}
	
\subfloat[Doppler error over the length axis of the vehicle]{
	\ifx\useExternalFigures\undefined
    \tikzsetnextfilename{ext_figures/trainingDataZxVd}%
    \input{figures/trainingDataZxVd.tikz}%

	\else
		\includegraphics{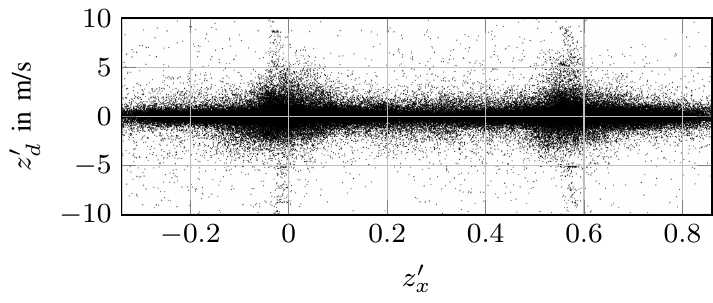}
	\fi
	\label{fig:trainingDataZxVd}}
\caption{Two views of the training data points}
\label{fig:trainingData}
\end{figure}

The entire data set comprises 336,287 data points from approximately 123 minutes of recorded sensor data. Two views of the data set are shown in \cref{fig:trainingData}. In particular, a top view of the measurements in normalized coordinates is shown in \cref{fig:trainingDataZxZy} and the Doppler error over the longitudinal axis in \cref{fig:trainingDataZxVd}. It is observable that most measurements originate from the vehicle surface and that deviations from the expected Doppler velocity 
mostly occur close to the front and rear axles.

There is an imbalance in the data set in terms of the number of measurements for different aspect angles. For example, it contains roughly three times more measurements from the rear perspective than from the front perspective. Also, there are about 20,000 data points in a \ang{5} interval around the rear perspective $x' = \ang{0}$, whereas the neighboring \ang{5} interval around $x' = \ang{-5}$ only contains 4736 data points. While the imbalance over distant aspect angles is mostly eliminated when computing the conditional density, local imbalances can introduce small biases. If only measurements from the rear surface are available for several time steps, for example, it was observed that a model that was learned from the entire data set tends to favor aspect angles around $x' = 0^\circ$. To avoid such issues, a balanced subset of data points was used as training set. It contains 95,688 data points that result in an even aspect angle histogram with \ang{5} bins.

\subsection{Resulting Variational Radar Model}
A modified MATLAB implementation\footnote{original implementation by Mo Chen,\\https://de.mathworks.com/matlabcentral/fileexchange/35362-variational-bayesian-inference-for-gaussian-mixture-model} of the \gls{VGM} was used to fit the mixture model to the training data. The number of components was set to $c = 70$ and the hyperparameter of the Dirichlet prior over the mixture weights was set to $\rho_0 = 1$. For the Gaussian-Wishart prior, the hyperparameters were set to $\beta_0 = 1$, $\nu_0 = |z_D| + 1$, $\gamma_0$ was set to the mean of all training points, and $\underline{V}_0$ was initialized as identity matrix. This results in a non-informative prior which does not assume a certain form of the \gls{VGM} parameters.

A useful feature of the \gls{VGM} approach is that it internally penalizes the model complexity. Unnecessary components---i.e. components which explain no or only very few measurements---automatically receive low weights. From the 70 initially proposed components, 20 received a mixing weight below $10^{-5}$. As these components do not contribute to the model and only increase computation time, they are removed. Thus, the number of components of the final mixture is $c = 50$.

\begin{figure}[!t]
	\centering
	\setlength\figurewidth{7.5cm}
	\ifx\useExternalFigures\undefined
    \tikzsetnextfilename{ext_figures/margDensity}%
    \input{figures/margDensity.tikz}%

	\else
		\includegraphics{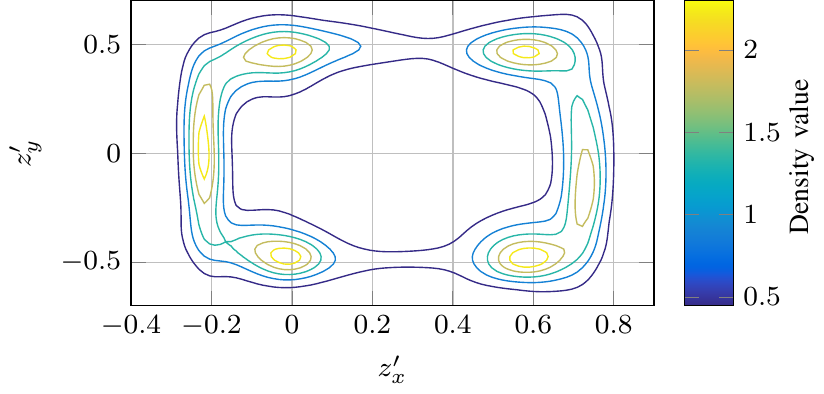}
	\fi
	\caption{Marginal density $p(z'_x, z'_y)$}
	\label{fig:margDensity}	
\end{figure}

As a visualization of the full, four-dimensional joint density $p(z', x')$ is difficult, \cref{fig:margDensity} illustrates the marginal density $p(z'_x, z'_y)$. The \gls{VGM} has identified that most measurements originate from the vehicle surface. Also, it identified the centers of the front and rear surface as well as the four wheels and wheel houses as typical measurement sources.

\begin{figure*}[!t]
\vspace{-0.3cm}
\centering
\hspace{-0.7cm}
\subfloat[$x' = -3$]{\setlength\figurewidth{\columnwidth}
	\ifx\useExternalFigures\undefined
    \tikzsetnextfilename{ext_figures/condDensity-3}%
    \input{figures/condDensity-3.tikz}%

	\else
		\includegraphics{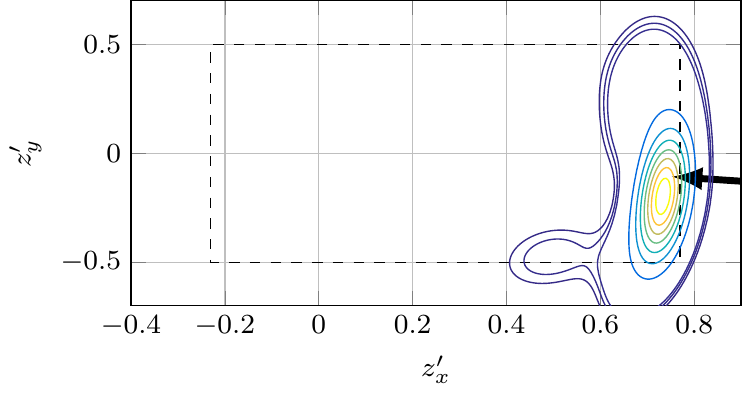}
	\fi
	\label{fig:condDensity-3}}%
\hspace{0.55cm}%
\subfloat[$x'= -\frac{\pi}{2}$]{\setlength\figurewidth{\columnwidth}
	\ifx\useExternalFigures\undefined
    \tikzsetnextfilename{ext_figures/condDensity-pi2}%
    \input{figures/condDensity-pi2.tikz}%

	\else
		\includegraphics{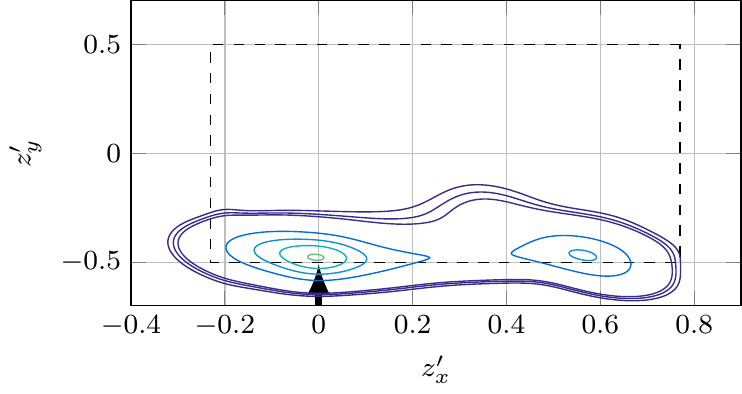}
	\fi
	\label{fig:condDensity-pi2}}%

\subfloat[$x' = -\frac{\pi}{4}$]{\setlength\figurewidth{\columnwidth}
	\ifx\useExternalFigures\undefined
    \tikzsetnextfilename{ext_figures/condDensity-pi4}%
    \input{figures/condDensity-pi4.tikz}%

	\else
		\includegraphics{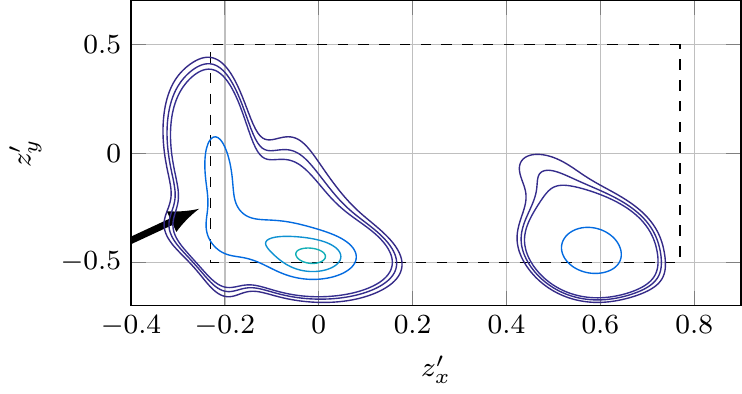}
	\fi
	\label{fig:condDensity-pi4}}%
\hspace{0.55cm}%
\subfloat[$x'= 0$]{\setlength\figurewidth{\columnwidth}
	\ifx\useExternalFigures\undefined
    \tikzsetnextfilename{ext_figures/condDensity0}%
    \input{figures/condDensity0.tikz}%

	\else
		\includegraphics{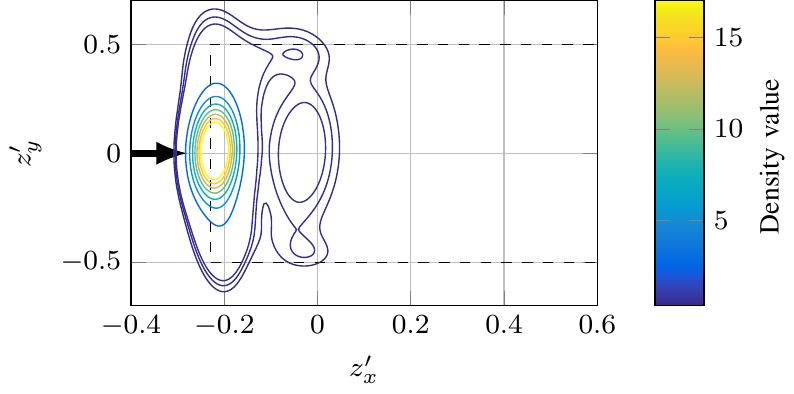}
	\fi
	\label{fig:condDensity0}}%
\caption{Marginal density $p(z'_x, z'_y | x')$ conditioned on the aspect angle $x'$. The line of sight between the sensor and the center of the rear axle is indicated by the arrow and the dashed rectangle depicts the normalized vehicle dimensions.}
\label{fig:condMargDensity}
\end{figure*}
The conditional density $p(z'_x, z'_y | x')$ shows where measurements are expected for a given aspect angle $x'$. \cref{fig:condMargDensity} depicts examples for different values of $x'$. \Cref{fig:condDensity-3} shows the conditional density when looking at the vehicle front. The aspect angle is close to the $\pi$, $-\pi$ boundary. Since the \gls{VGM} does not consider the periodic nature of the aspect angle, an abrupt change in the involved mixture components occurs when the sign of the aspect angle changes. This could be further improved by adapting the standard \gls{VGM} to periodic states. As the components on both sides are similar and expand over the boundary, however, this issue has so far not been noticeable during application. A view from the right side of the vehicle is illustrated in \cref{fig:condDensity-pi2}. Clearly, measurements are expected close to the right vehicle surface ($z'_y = -0.5$). Also, the positions of the right wheels are identifiable as frequent measurement sources. This effect is again visible in \cref{fig:condDensity-pi4} where the vehicle is viewed from rear right. In addition to the wheels, the vehicle corner becomes another prominent feature. When viewed from the rear, measurements tend to originate from the center of the rear surface as shown in \cref{fig:condDensity0}. Additionally, relatively low weighted components on the vehicle interior come into play. A possible explanation for the components close to $z'_x = 0$ is that the sensor receives reflections from the rear axle or the edge of the vehicle roof.

\begin{figure}[!t]
	\centering
	\setlength\figurewidth{7.5cm}
	\ifx\useExternalFigures\undefined
    \tikzsetnextfilename{ext_figures/dopplerDensity}%
    \input{figures/dopplerDensity.tikz}%

	\else
		\includegraphics{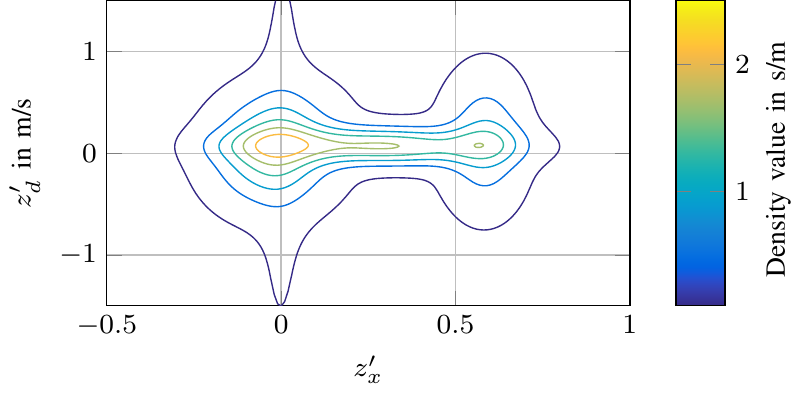}
	\fi
	\caption{Conditional density $p(z'_x, z'_d | z_y = -0.5, x' = -\frac{\pi}{2})$}
	\label{fig:condDensityDoppler}
\end{figure}

The conditional density $p(z'_x, z'_d | z_y = -0.5, x' = -\frac{\pi}{2})$ is depicted in \cref{fig:condDensityDoppler}. It shows the density of the Doppler error on the right vehicle surface over the vehicle length when looking from the right. It can be observed that the model expects larger Doppler errors in the vicinity of the wheels. In this case, the \gls{VGM} has automatically learned the occurence of spurious measurements from rotating wheels.

\section{Tracking Using Experimental Data}\label{s:evaluation}
In this section, results for the tracking performance of the multi-object tracking approach in combination with the variational radar model are presented. The algorithm was implemented in MATLAB and applied to different experimental scenarios that were recorded using the same ego-vehicle as in section \cref{s:expRadarModel}. The section starts with some practical remarks on the implementation and the tracking accuracy is subsequently assessed for single and multi-vehicle scenarios. The performance is compared to the manually designed direct scattering approach from \cite{Scheel.2016a}. It differs in the single object likelihood and uses a clutter density which is defined in polar coordinates. The multi-object filter core is identical.

\subsection{Practical Implementation Issues}

\subsubsection{Number of Particles}
Upon initialization, the number of particles for representing the birth density $p_B(x, \ell)$ is 900 to cover the wide range of possible states. This number is gradually reduced by 100 in the following update steps until the number of particles reaches 300, which is the steady state value for tracked objects.

\subsubsection{Constraints on Vehicle Dimensions}
The dimensions of a vehicle are restricted to maximum and minimum values. These are $a_\text{min} = \SI{1.4}{m}$ and $a_\text{max} = \SI{2.5}{m}$ for the width as well as $b_\text{min} = \SI{2.5}{m}$ and $b_\text{max} = \SI{7}{m}$ for the length. Additionally, the ratio between the length and width is restricted to minimum and maximum values of 1.7 and 3.5, respectively. Thus, only extent hypotheses with reasonable proportions are allowed.

\subsubsection{Process and Measurement Model Parameters}
During prediction, process noise is added to the kinematic states of the vehicles. The noise is modeled as uniform distributions centered at 0 and sampled for each particle. The maximum values are defined for the interval of one second and adjusted proportionally to the time difference between consecutive prediction steps. The normalized values are \SI{3}{m/s} for the position, \SI{0.698}{rad/s} for the angle, \SI{9}{m/s^2} for the velocity, and \SI{3}{rad/s^2} for the yaw rate. The probability of persistence is made dependent on the time between two consecutive updates and determined from an exponential distribution which models that an object persists for an average of \SI{10}{s} in and \SI{0.1}{s} outside the \gls{FOV}. In the measurement model, the probability of detection is set to 0.8 and slowly decreased towards the boundaries of the \gls{FOV}. The expected number of object and clutter measurements are set to $\lambda_T = 5$ and $\lambda_C = 30$. 

\subsubsection{Partitioning and Association}
Evaluating all possible measurement partitions and cluster-to-object associations as demanded by the multi-object likelihood function \cref{eq:moLikelihood} is computationally intractable even for a moderate amount of measurements. Therefore, only meaningful partitions are evaluated for obtaining the posterior multi-object density \cref{eq:posteriorDensity}. Partitions are generated in two ways. In a first step, DBSCAN \cite{Ester.1996} with different distance thresholds between \SI{0.5}{m} and \SI{5}{m} is applied. Additionally, the predicted tracks are used to generate partitions by combining all measurements that are in the vicinity of an existing track. The resulting clusters in the partitions and particularly the contained Doppler measurements are further analyzed. If the measurements do not conform to consistent rigid body motion, the clusters are split and additional partitions with the resulting subclusters are added. This allows excluding clutter measurements as for example measurements from rotating wheels. As will be shown, this step is mainly necessary for the direct scattering tracking approach that is used for comparison. For each multi-object state hypothesis, the ten best association variants are determined using Murty's algorithm \cite{Murty.1968} and evaluated.

\subsubsection{Initialization and Pruning}
New vehicle hypotheses are initialized as soon as there is a measurement cluster with at least two measurements that exhibit relevant Doppler velocities and have not considerably contributed to updating an existing vehicle. The goal is to avoid the creation of new hypotheses for stationary objects or from single temporary clutter measurements. New vehicle hypotheses are assigned a birth existence probability of $r_B^{(\ell)} = 0.1$ and the birth density $p_B(x, \ell)$ is formed by creating particles with different plausible states. For poses where the length is observable, suitable length hypotheses between \SI{2.5}{m} and \SI{7}{m} are initialized. If the length is not observable, random values between \SI{4}{m} and \SI{5}{m}, that are closer to typical vehicle lengths, are created. As soon as the existence probability of a vehicle hypothesis falls below 0.01, it is pruned from the multi-object density.

\subsubsection{Ego-Motion Compensation}
Motion of the ego-vehicle affects tracking in two ways and hence, compensation procedures are added. In measurement processing, the contribution of ego-vehicle motion is computed and removed from the Doppler measurements. Also, the prediction routine needs to account for the moving ego-vehicle coordinate system in which the vehicles are tracked. Therefore, an additional step transforms the vehicles from the last to the current vehicle coordinate system.

\subsubsection{Sensor Fusion}
The presented tracking approach is used in a centralized fusion architecture to fuse the data from all four radar sensors of the vehicle. That is, a new update is triggered each time new data from a sensor arrives and the information is fused into the posterior multi-object density. Measurements arrive in order of recording time and out-of-sequence problems are not considered.

\subsubsection{Computation Time}
The prototype MATLAB implementation presented in this paper is not intended for real-time calculations and the runtime is therefore not in the focus of the evaluations. A rough analysis has, however, revealed that approximately 85\% of the computation time is spent on evaluating the variational radar model. Also, first porting to C++ on a single core has indicated potential for object update times in the two-digit milliseconds range. To achieve real-time capability, the crucial point of future work is thus to find fast implementations for evaluating the densities and to use parallelization for the particle implementation.

\subsection{Single-Object Accuracy}
The tracking accuracy for a single vehicle is evaluated on experimental data that was recorded using the same ego and target vehicle on a different day. In total, ten scenarios were evaluated. They comprise situations with oncoming and crossing traffic or passing and turning vehicles. Due to the Monte Carlo implementation, which involves random generation and propagation of particles, estimation results are subject to random effects. To diminish these effects, all scenarios were evaluated \nRadarRuns{} times and the results are averaged over these runs.

\begin{figure*}[ht!]
\centering
\setlength\figureheight{10cm}
\setlength\figurewidth{16cm}
\ifx\useExternalFigures\undefined
    \tikzsetnextfilename{ext_figures/vmEightAccuracy/vmEightAccuracy}%
    \input{figures/vmEightAccuracy/vmEightAccuracy.tikz}%

\else
	\includegraphics{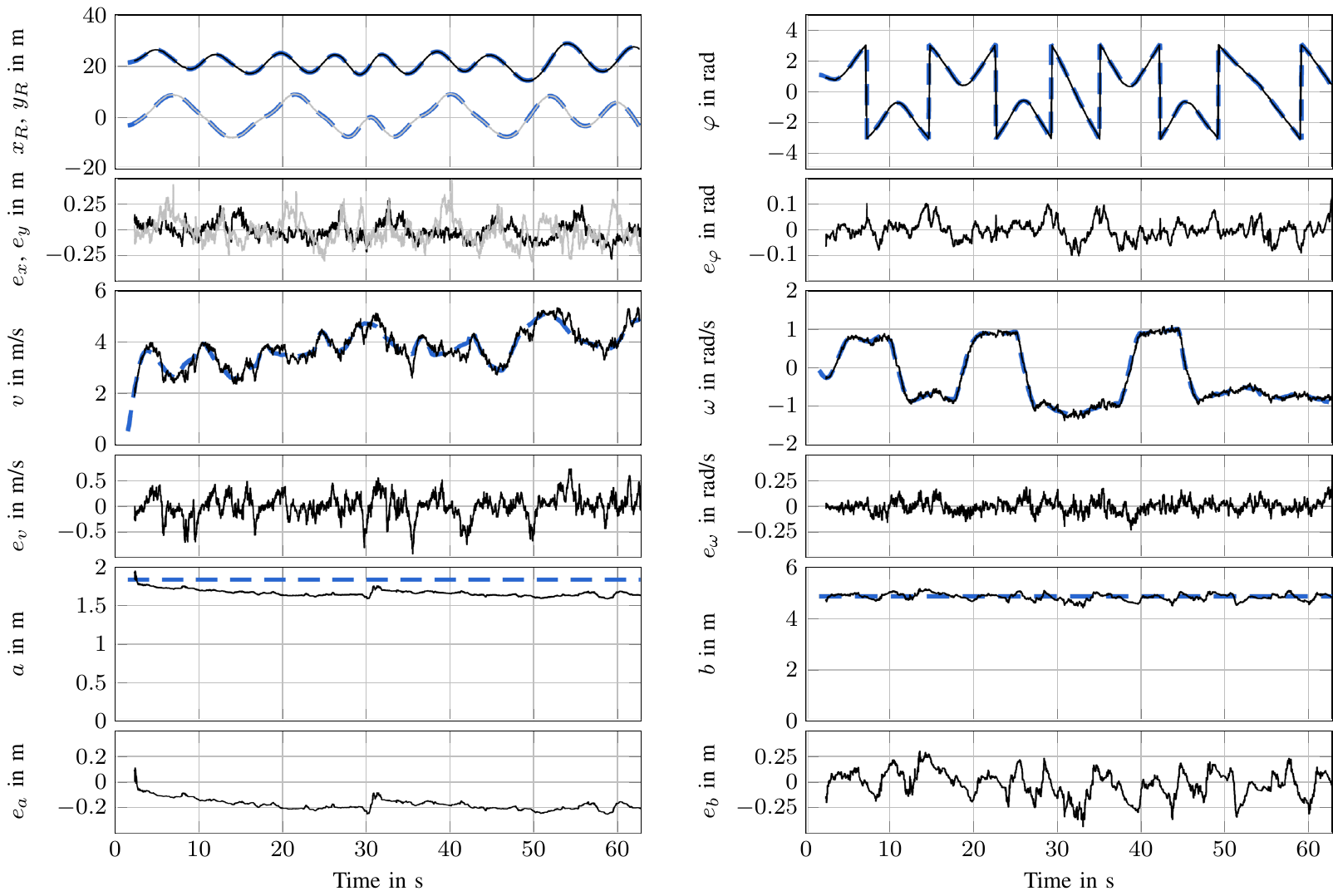}
\fi
\caption{Figure eight scenario: Estimates (solid) and ground truth (dashed) as well as errors $e$ averaged over \nRadarRuns{} runs. The y-position is plotted in gray.}
\label{fig:accuracy}
\end{figure*}

\begin{figure}
\centering
\setlength\figurewidth{\columnwidth}
\ifx\useExternalFigures\undefined
    \tikzsetnextfilename{ext_figures/eightExcerpt}%
    \input{figures/eightExcerpt.tikz}%

\else
	\includegraphics{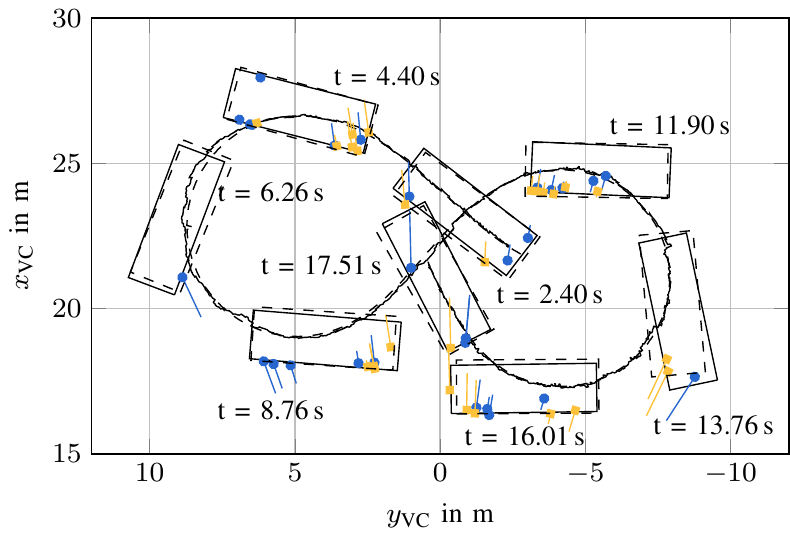}
\fi
\caption{Excerpt of the figure eight scenario: radar measurements with indicated Doppler velocity from the front left (\protect\includegraphics{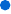}) and front right (\protect\includegraphics{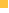}) sensor, estimated trajectory (solid) and exemplary vehicle poses (solid rectangles), reference trajectory (dashed) and reference poses (dashed rectangles)}
\label{fig:eightExcerpts}
\end{figure}

In the following, one scenario in which the target vehicle drives a figure eight in front of the stationary ego-vehicle and is visible to the two front sensors is examined in detail. The scenario is challenging for several reasons: The aspect angle on the target vehicle changes constantly, it deviates from classical longitudinal traffic scenarios in that it contains a turning vehicle and cross traffic where the Doppler measurements do generally not equal the vehicle speed, and it is highly dynamic with yaw rates up to \SI{60}{\degree /s}. \Cref{fig:accuracy} shows the estimation results, reference values, and resulting estimation errors for all components of the state vector. An excerpt of the scenario from an exemplary run is shown in \cref{fig:eightExcerpts}.

\Cref{tab:rmseValues} lists \gls{RMSE} values for the figure eight scenario as well as combined values over all single object scenarios. For comparison, results for the direct scattering approach are also provided. The variational radar model considerably outperforms the manually designed direct scattering model for all states. Despite the complicated maneuver, it achieves especially precise estimation results for the figure eight scenario. The accuracy decreases for both approaches when averaging over all scenarios. In contrast to the figure eight scenario, where the target vehicle is visible from all four sides, it is only partially visible over longer periods of time in other scenarios. This deteriorates size estimation and leads to correlated position errors. Also, vehicles in greater distance or vehicles with straight motion trend to yield fewer measurements. Thus, accurate orientation estimation is more difficult. An exemplary case is presented in the next section.
\begin{table}
\centering
\caption{RMSE values for the figure eight scenario (8) and all single object scenarios (all) using the the variational model (VM) and the direct scattering model (DSM)}
\label{tab:rmseValues}
\begin{tabular}{lccccc}
\toprule
States & VM 8 & DSM 8 & VM all & DSM all\\
\midrule
$x_R$ in \si{m}             & 0.10 & 0.25 & 0.27 & 0.31 \\
$y_R$ in \si{m}             & 0.13 & 0.19 & 0.26 & 0.40 \\
$\varphi$ in \si{\degree}   & 2.29 & 3.60 & 7.28 & 9.41 \\
$v$ in \si{m}               & 0.25 & 0.35 & 0.36 & 0.55 \\
$\omega$ in \si{\degree /s} & 3.57 & 6.15 & 5.54 & 8.63 \\
$a$ in \si{m}               & 0.19 & 0.33 & 0.26 & 0.32 \\
$b$ in \si{m}               & 0.16 & 0.49 & 0.28 & 0.50 \\
\bottomrule
\end{tabular}
\end{table}

\subsection{Multi-Object Performance}
The multi-object performance is assessed using nine different scenarios with three vehicles: the ego-vehicle, the E-Class target vehicle, and an additional Mercedes C-Class station wagon (S205), which is also equipped with a \gls{DGPS}/\gls{IMU} system. The nine scenarios comprise different situations such as oncoming traffic, cross traffic, overtaking, and occlusions. Again, the results are averaged over \nRadarRuns{} Monte Carlo runs.

\begin{figure*}
\centering
\setlength\figurewidth{2\columnwidth}
\ifx\useExternalFigures\undefined
    \tikzsetnextfilename{ext_figures/oncoming}%
    \input{figures/oncoming.tikz}%

\else
	\includegraphics{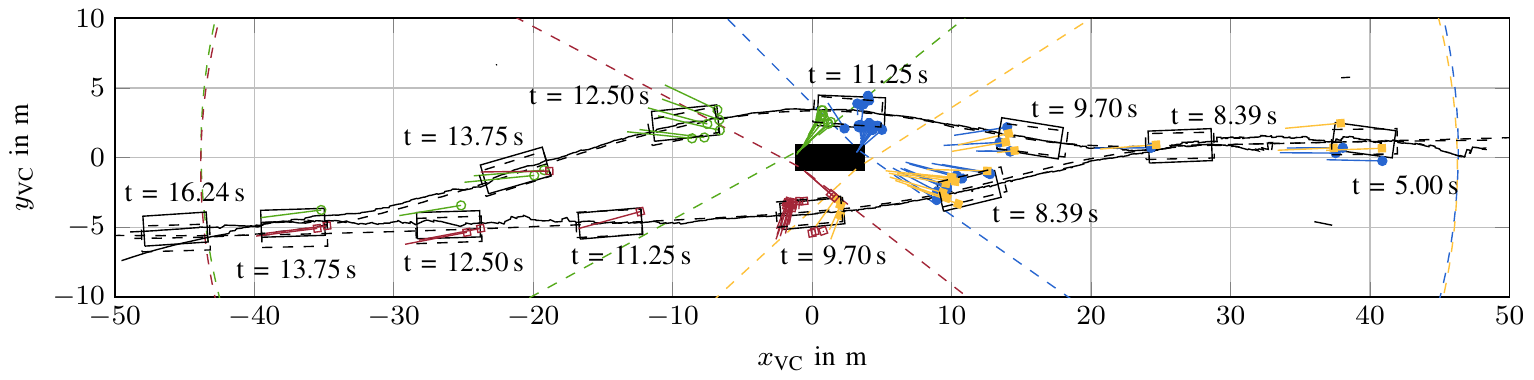}
\fi
\caption{Scenario with two oncoming vehicles: Estimated (solid) and ground truth (dashed) trajectories, exemplary vehicle poses (estimates: solid rectangles, ground truth: dashed rectangles), corresponding measurements with Doppler velocity (front left: \protect\includegraphics{ext_figures/blueDot.pdf}, front right: \protect\includegraphics{ext_figures/yellowSquare.pdf}, rear left: \protect\includegraphics{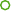}, rear right: \protect\includegraphics{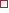}), and sensor \glspl{FOV}}
\label{fig:oncoming}
\end{figure*}

\begin{figure}
\centering
\setlength\figurewidth{0.9\columnwidth}
\setlength\figureheight{1.5cm}
\ifx\useExternalFigures\undefined
    \tikzsetnextfilename{ext_figures/oncomingCard}%
    \input{figures/oncomingCard.tikz}%

\else
	\includegraphics{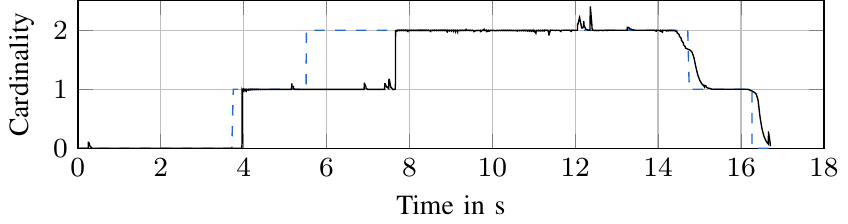}
\fi
\caption{Cardinality estimate (black) and ground truth (dashed) for the scenario with two oncoming vehicles}
\label{fig:oncomingCard}
\end{figure}

An exemplary run of one of the scenarios is shown in \cref{fig:oncoming}. Here, two vehicles are approaching the stationary ego-vehicle and pass it on both sides. The target vehicles are continuously tracked and cross the \glspl{FOV} of all four radar sensors. As mentioned before, it is observable that the tracking results are very precise in the direct vicinity of the ego-vehicle and become less precise towards \gls{FOV} boundaries as measurements become more scarce and less accurate. The cardinality estimate is plotted in \cref{fig:oncomingCard}. As soon as the vehicles enter the \gls{FOV}, the true cardinality rises to one and then two. It decreases once the vehicles leave the \gls{FOV}. The filter is mostly able to correctly estimate the cardinality. Yet, it takes a considerable amount of time to initialize the second track. This is because the second vehicle only creates single measurements in the far range while the initialization routine expects at least a cluster of two.
Thus, the vehicle is not set up before two measurements are created at a distance of approximately \SI{35}{m}. An adaption of the initialization routine could eliminate this issue.

\begin{figure}
\centering
\setlength\figurewidth{0.45\columnwidth}
\subfloat[Direct scattering model]{
	\ifx\useExternalFigures\undefined
    \tikzsetnextfilename{ext_figures/wheelMeasDs}%
    \input{figures/wheelMeasDs.tikz}%

	\else
		\includegraphics{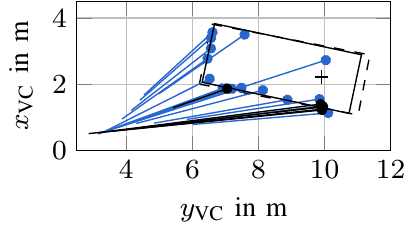}
	\fi	
	\label{fig:wheelMeasDs}}
\subfloat[Variational radar model]{
	\ifx\useExternalFigures\undefined
    \tikzsetnextfilename{ext_figures/wheelMeasVm}%
    \input{figures/wheelMeasVm.tikz}%

	\else
		\includegraphics{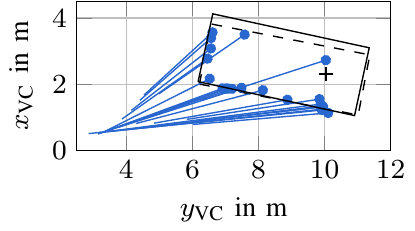}
	\fi
	\label{fig:wheelMeasVm}}
\caption{Comparison of the measurement clusters that contributed the most during update: cluster measurements (\protect\includegraphics{ext_figures/blueDot.pdf}), other measurements (\protect\includegraphics{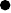}), average vehicle estimate with center of the rear axle (rectangle and cross), reference vehicle (dashed rectangle)}
\label{fig:wheelMeas}
\end{figure}

\Cref{fig:wheelMeas} depicts two additional excerpts of the scenario. Here, the average estimate of the upper vehicle at \SI{10.54}{s} and corresponding radar measurements from the front left sensor are shown for direct scattering model (\cref{fig:wheelMeasDs}) and for the variational radar model (\cref{fig:wheelMeasVm}). Additionally, the measurements from the cluster that has contributed the most to updating the vehicle are indicated. The direct scattering model favors a cluster which excludes six measurements that originate from the wheels and exhibit an especially large or small Doppler velocity. This due to the fact that measurements from rotating wheels are not considered in the model. In this case, the direct scattering model profits from the multi-object approach which allows for different partitioning hypotheses. In contrast, the variational model has learned the effect of spurious measurements of the rotating wheels and uses a cluster which contains all measurements. The additional information helps to locate the position of the vehicle axes and might thus be a cause for the improved length estimation performance.

\begin{figure}[!t]
	\centering
	\setlength\figurewidth{7.5cm}
	\ifx\useExternalFigures\undefined
    \tikzsetnextfilename{ext_figures/twoParallel}%
    \input{figures/twoParallel.tikz}%

	\else
		\includegraphics{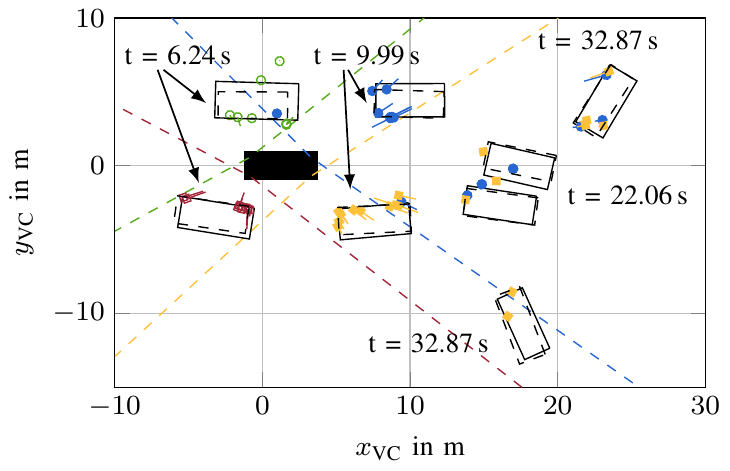}
	\fi
	\caption{Two vehicles driving closely: Estimated (solid) and ground truth (dashed) vehicle poses, corresponding measurements with Doppler velocity (front left: \protect\includegraphics{ext_figures/blueDot.pdf}, front right: \protect\includegraphics{ext_figures/yellowSquare.pdf}, rear left: \protect\includegraphics{ext_figures/greenCircle.pdf}, rear right: \protect\includegraphics{ext_figures/redSquare.pdf}), and sensor \glspl{FOV}}
	\label{fig:twoParallel}	
\end{figure}

A second multi-object scenario is shown in \cref{fig:twoParallel}. The ego-vehicle is first passed by the two target vehicles on both sides. Then, they closely drive in parallel in front of the ego-vehicle before they depart to the left and the right at around 30 seconds. Since the ego-vehicle is moving, the trajectories, which are estimated in vehicle coordinates, are difficult to visualize and thus not plotted. At \SI{22.06}{s}, the benefit of using a multiple extended object tracking approach which considers multiple partitions and associations becomes apparent: The target vehicles are driving so close to each other that the distance between measurements from both vehicles is lower as for example the distance in between measurements from the upper vehicle at \SI{32.87}{s}. Using a clustering routine with fixed parameter set would not be effective in both situations and classical preprocessing routines would most likely merge the two close-by vehicles into a single object. By considering different hypotheses, however, the algorithm is able to find the right associations in both cases.

\begin{figure}[t]
\centering
\setlength\figurewidth{0.9\columnwidth}
\setlength\figureheight{2cm}
\ifx\useExternalFigures\undefined
    \tikzsetnextfilename{ext_figures/cardErrorHist}%
    \input{figures/cardErrorHist.tikz}%

\else
	\includegraphics{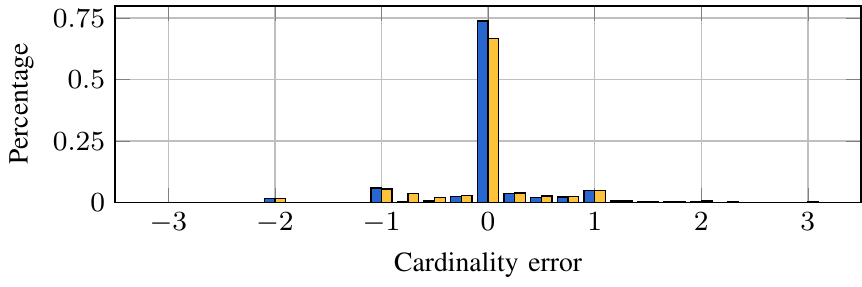}
\fi
\caption{Histogram of cardinality estimation errors for the multi-object scenarios: variational model (\protect\includegraphics{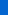}) and direct scattering model (\protect\includegraphics{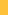})}
\label{fig:cardErrorHist}
\end{figure}

The cardinality estimates of the variational and the direct scattering approach are compared using histograms of cardinality errors over all nine multi-object scenarios. To obtain the ground truth, all vehicles in the sensor \gls{FOV} with a speed greater than 1 m/s were counted. The histograms are shown in \cref{fig:cardErrorHist}. While the direct scattering approach estimates the correct cardinality in \correctCardDS{} of the update steps, the variational approach is correct in \correctCardVM{} of the time. The direct scattering approach overestimates the cardinality in \overestCardDS{} of the update steps, whereas the percentage is reduced to \overestCardVM{} when using the variational radar model. This suggests that the variational radar model performs better in distinguishing clutter from actual vehicles. False tracks are mostly caused by spurious measurements with non-zero Doppler velocity and may survive if there are matching clutter measurements from stationary objects over several time steps. Such cardinality errors are caused by a violation of the assumption of independent and uniform clutter. The cardinality is underestimated in \underestCardDS{} of the update steps for the direct scattering model and \underestCardVM{} for the variational radar model. In these cases, the filter has either not yet initialized a track, has assigned too low existence probabilities to the vehicle hypotheses, or vehicle tracks are temporarily lost. Delayed initialization makes up \cardErrorLateInitDS{} and \cardErrorLateInitVM{} of the cardinality errors for the direct scattering and variational radar models, respectively.

\subsection{Generalization}
So far, the variational model was tested using the same E-Class target vehicle that generated the training data and a rather similar C-Class vehicle. To demonstrate that the model is applicable to a wider range of vehicle types, it was applied to an urban scenario. The ego-vehicle stands at a T-intersection, while eleven different vehicles pass it. The vehicle types range from compact cars over sedans and convertibles to vans. Unfortunately, no accurate ground truth is available for these vehicles. Therefore, the vehicle poses and dimensions where manually labeled using the lidar sensor of the ego-vehicle as reference. The labels are only available in the lidar \gls{FOV} which approximately covers \ang{100} in front of the ego-vehicle.

\begin{figure*}
\centering
\setlength\figurewidth{\columnwidth}
\subfloat[Occlusion situation]{
	\ifx\useExternalFigures\undefined
    \tikzsetnextfilename{ext_figures/generalization2}%
    \input{figures/generalization2.tikz}%

	\else
		\includegraphics{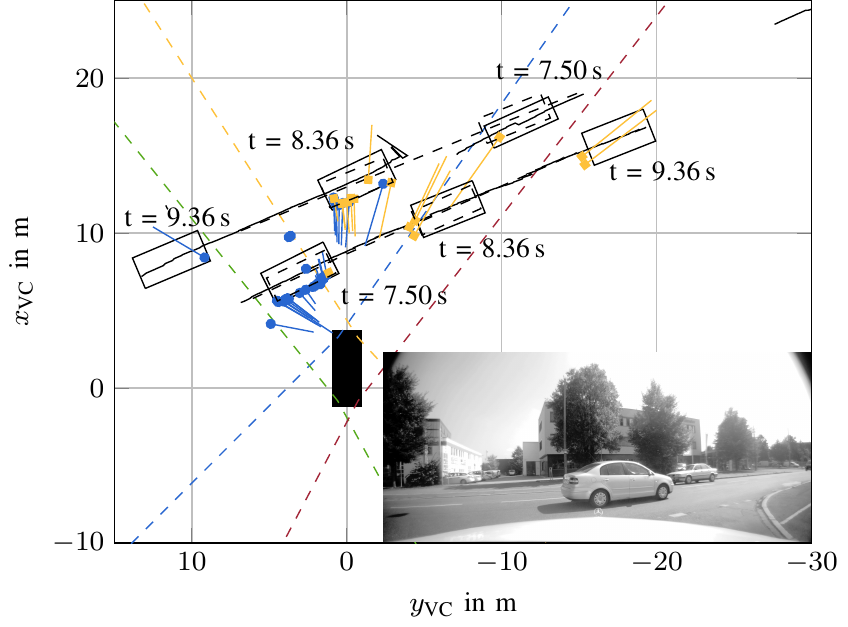}
	\fi
	\label{fig:generalization2}}
\subfloat[Turning van and two other vehicles]{
	\ifx\useExternalFigures\undefined
    \tikzsetnextfilename{ext_figures/generalization3}%
    \input{figures/generalization3.tikz}%

	\else
		\includegraphics{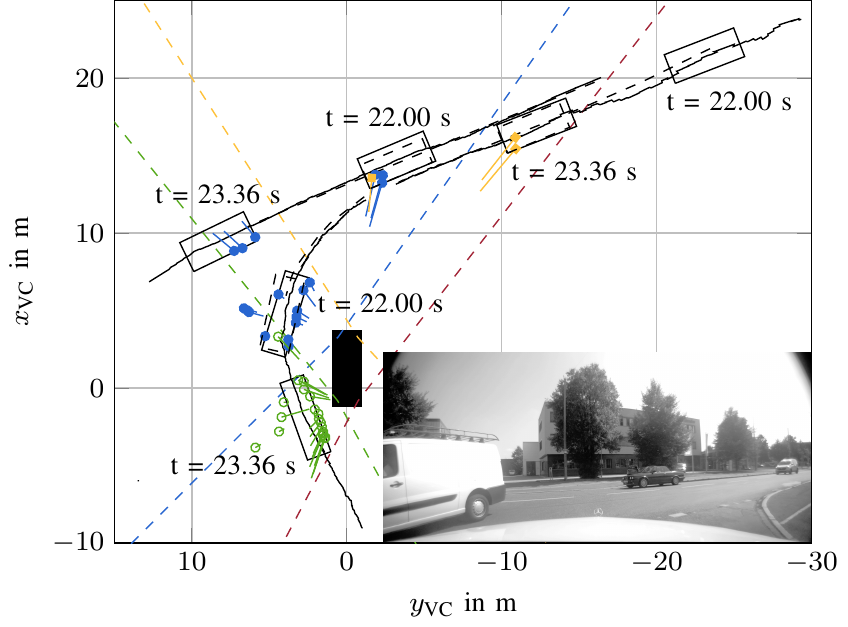}
	\fi
	\label{fig:generalization3}}
\caption{T-intersection scenario: Two excerpts with estimated (solid) and true (dashed) trajectories (only available in the lidar \gls{FOV}), exemplary vehicle poses (solid rectangles) and true poses (dashed rectangles), corresponding measurements (front left: \protect\includegraphics{ext_figures/blueDot.pdf}, front right: \protect\includegraphics{ext_figures/yellowSquare.pdf}, rear left: \protect\includegraphics{ext_figures/greenCircle.pdf}, rear right: \protect\includegraphics{ext_figures/redSquare.pdf}), and sensor \glspl{FOV}}
\label{fig:generalization}
\end{figure*}

Two exemplary situations are shown in \cref{fig:generalization}. \Cref{fig:generalization2} shows one of the most challenging situations for the algorithm. Here, two sedan vehicles cross in front of the ego-vehicle and the front sedan temporarily occludes the second vehicle. While the front sedan is tracked continuously, the track of the rear sedan is lost during occlusion. In this situation, the sensors do not provide measurements from this vehicle over a period of 25 update steps. This causes the probability of existence to drop below the pruning threshold. Once the vehicle is visible again, it is reinitialized. Dropping the assumption of object measurements being generated independently and including an occlusion model as for example used in \cite{Granstroem.2014} could alleviate this issue.
\Cref{fig:generalization3} shows a constellation of three vehicles. A van takes a left turn past the ego-vehicle whereas a convertible and a compact van are driving straight. Here, all three vehicles are continuously tracked.

All in all, the variational radar model did not show difficulties with a particular vehicle type even though it was trained using data from a single vehicle. The \gls{RMSE} values where computed with respect to the manually created labels in front of the ego-vehicle and averaged over the eleven vehicles and \nRadarRuns{} Monte Carlo runs. They are \tIntVMx{} and \tIntVMy{} for $x_R$ and $y_R$, \tIntVMPhi{} for $\varphi$, \tIntVMa{} for the width and \tIntVMb{} for the length. Track estimates are available for the labeled vehicles in \tIntVMTrackAvail{} of the update steps. The remaining 4.9\% are due to delayed initialization of entering vehicles and the track loss during occlusion (cf. \cref{fig:generalization2}). From visual inspection, an expectable degradation of performance occurs in the far field where the number as well as the accuracy of measurements decreases and the number of misdetections and clutter increases.

The indicated ability to generalize to other vehicles is not surprising as even high-resolution radar measurements are not yet at the resolution performance of other sensor types such as lidar. Hence, the rough extent of the vehicles is observable but the details that distinguish different vehicles are still concealed. Problems are expected as soon as the vehicle appearance changes drastically, e.g. with additional wheels or truck bodies with distinct reflection characteristics.


\section{Conclusion}\label{s:conclusion}
In this paper, a variational radar model for vehicles that is learned from actual radar data was presented and included in a \gls{FISST}-based multi-object filter. Both measurement modeling and multi-object filtering are formulated in an integral Bayesian fashion. \gls{FISST} provides a rigorous mathematical formulation of the multi-object problem which allows for a natural incorporation of the variational radar model. The multi-object filter considers object dependencies, e.g. that objects should not overlap and that measurements may only originate from one object, and is able to filter over several measurement associations and partitions.

By learning a vehicle model from actual radar data, the variational radar model is a close approximation of the true measurement likelihood and avoids the need for excessive manual engineering. Also, it was shown that it is able to outperform state of the art extended object methods in both the single and multi-object performance. The capability to generalize to objects that are not contained in the training data was shown using a real-world example.

There are several possible extensions of the approach to overcome some limitations. For example, using other nonlinear estimation techniques such as a \gls{UKF} for the single-object densities could simplify the approach and facilitate fast real time implementations. So far, the approach does not learn all parameters that are involved in the multi-object likelihood. Learning additional parameters such as the expected number of measurements or the clutter densities could further improve modeling accuracy. Studies on reduction of the training data, using alternative dimension reduction techniques, exploiting vehicle symmetries, approximation of the resulting Student's~t mixtures by simpler Gaussian mixtures, and ablation studies could provide further insight in simplifications of the approach. Also, the \gls{VGM} approach could be transferred to other object classes to eventually achieve a multi-class tracking framework that is able to handle all types of traffic participants.

\ifCLASSOPTIONcaptionsoff
  \newpage
\fi



\bibliographystyle{IEEEtran.bst}
\bibliography{IEEEtranControl,bibliography}

%
%

%

\begin{IEEEbiography}[{\includegraphics[keepaspectratio]{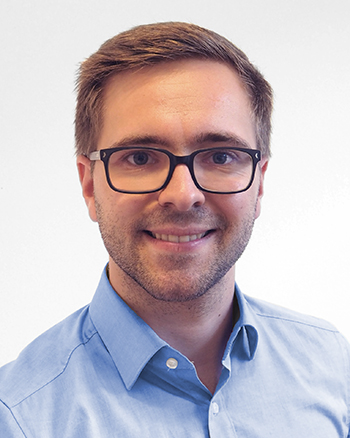}}]{Alexander Scheel}
was born in Heilbronn, Germany, in 1987. In 2013, he received his Diploma degree (equivalent to M.Sc. degree) in automotive and engine technology from the University of Stuttgart. Since December 2013, he has been working as research assistant at the Institute for Measurement, Control, and Microtechnology at Ulm University where he became group leader for autonomous driving in 2016. In 2018, he joined the autonomous driving department of Robert Bosch GmbH. Alexander Scheel specializes in environment perception for autonomous vehicles. In particular, his research interests include multi-object tracking, sensor data fusion, and measurement models for extended objects.
\end{IEEEbiography}

\begin{IEEEbiography}[{\includegraphics[keepaspectratio]{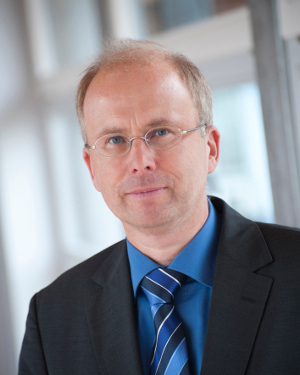}}]{Klaus Dietmayer}
(M’05) was born in Celle, Germany in 1962. He received his Diploma degree (equivalent to M.Sc. degree) in 1989 electrical engineering from the Technical University of Braunschweig (Germany), and the Dr.-Ing. degree (equivalent to Ph.D.) in 1994 from the University of Armed Forces in Hamburg (Germany). In 1994 he joined the Philips Semiconductors Systems Laboratory in Hamburg, Germany as a research engineer. Since 1996 he became a manager in the field of networks and sensors for automotive applications. In 2000 he was appointed to a professorship at Ulm University in the field
of measurement and control. Currently he is Full Professor and Director of the Institute of Measurement, Control and Microtechnology in the school of Engineering and Computer Science at Ulm University. His research interests include information fusion, multi-object tracking, environment perception for advanced automotive driver assistance, and E-Mobility. Klaus Dietmayer is member of the IEEE and the German society of engineers VDI/VDE.
\end{IEEEbiography}






\end{document}